\documentclass[a4paper,fleqn,usenatbib]{mnras}
\usepackage[T1]{fontenc}
\usepackage{ae,aecompl}
\usepackage{longtable}
\usepackage{makecell}
\usepackage{multirow}
\usepackage{supertabular,booktabs}
\usepackage{amsmath}
\usepackage{upgreek}

\usepackage{graphicx}	% Including figure files
\usepackage{amssymb}	% Extra maths symbols

\usepackage{url}
\usepackage{color}
\usepackage{url}
\usepackage{journals}
\usepackage{array,booktabs,arydshln,xcolor}
\usepackage{colortbl}
\usepackage{xcolor}
\usepackage{multicol}

\definecolor{dblue}{rgb}{ 0.08235294,  0.39607843,  0.75294118}
\newcommand{\msun}{{\rm M}_{\odot}}
\newcommand{\ergs}{erg\,s$^{-1}$}

\newcommand {\be}{\begin {equation}}
\newcommand {\ee}{\end {equation}}

\newcommand{\jttt}{J1023}
\newcommand{\fgl}{J1544}

\newcommand{\lr}{$L_\mathrm{R}$}
\newcommand{\lx}{$L_\mathrm{X}$}

\usepackage{soul}

\title[Simultaneous radio and X-ray observations of \fgl]
      {Simultaneous radio and X-ray observations of the transitional millisecond pulsar candidate 3FGL~J1544.6$-$1125}
 \author[Gusinskaia et al.]
    {N.~V. Gusinskaia,$^{1,2,3,4,5}$\thanks{E-mail: gusinskaia@astro.utoronto.ca}
    A.~D. Jaodand,$^{8,14}$
    J.~W.~T. Hessels,$^{1,2}$
    S. Bogdanov,$^{6}$
\newauthor A.~T. Deller,$^{7}$
    J.~C.~A. Miller-Jones,$^{9}$
    T.~D. Russell,$^{1,10}$ 
    A. Patruno,$^{11,12,2}$ \&
\newauthor A.~M. Archibald$^{1,13}$\\
$^1$Anton Pannekoek Institute for Astronomy, University of Amsterdam, Science Park 904, 1098 XH Amsterdam, The Netherlands\\
$^2$ASTRON, the Netherlands Institute for Radio Astronomy, Postbus 2, 7990 AA, Dwingeloo, The Netherlands\\
$^{3}$Department of Astronomy and Astrophysics, University of Toronto, 50 St. George Street, Toronto, ON M5S 3H4, Canada\\
$^{4}$Canadian Institute for Theoretical Astrophysics, University of Toronto, 60 St. George Street, Toronto, ON M5S 3H8, Canada\\
$^{5}$Dunlap Institute for Astronomy \& Astrophysics, University of Toronto, 50 St.~George Street, Toronto, ON M5S 3H4, Canada\\
$^6$Columbia Astrophysics Laboratory, Columbia University, 550 West 120th Street, New York, NY, 10027, USA\\
$^7$Centre for Astrophysics and Supercomputing, Swinburne University of Technology, P.O. Box 218, Hawthorn, VIC 3122, Australia\\
$^8$Cahill Center for Astrophysics, 1216 E. California Blvd, California Institute of Technology, Pasadena, CA 91125, USA\\
$^9$International Centre for Radio Astronomy Research, Curtin University, GPO Box U1987, Perth, WA 6845, Australia\\
$^{10}$INAF-Istituto di Astrofisica Spaziale e Fisica Cosmica, Via U. La Malfa 153, I-90146 Palermo, Italy\\
$^{11}$Institute of Space Sciences (IEEC–CSIC), Carrer de Can Magrans s/n, E-08193 Barcelona, Spain\\
$^{12}$Leiden Observatory, Leiden University, Neils Bohrweg 2, 2333 CA Leiden, The Netherlands\\
$^{13}$Newcastle University, NE1 7RU, UK\\
$^{14}$Center for Astrophysics, Harvard \& Smithsonian, 60 Garden St., Cambridge, MA 02138, USA}
\date{Accepted xxxx xxx xx.  Received xxx xxxx xx; in original form 2023}

\pubyear{2023}

\begin{document}
\label{firstpage}
\pagerange{\pageref{firstpage}--\pageref{lastpage}}
\maketitle

\begin{abstract}

Transitional millisecond pulsars (tMSPs) are neutron-star systems that alternate between a rotation-powered radio millisecond pulsar state and an accretion-disk-dominated low-mass X-ray binary (LMXB)-like state on multi-year timescales. During the LMXB-like state, the X-ray emission from tMSPs switches between `low' and `high' X-ray brightness modes on a timescale of seconds to minutes (or longer), while the radio emission shows variability on timescales of roughly minutes. Coordinated VLA and {\it Chandra} observations of the nearby tMSP PSR~J1023+0038 uncovered a clear anti-correlation between radio and X-ray luminosities such that the radio emission consistently peaks during the X-ray low modes. In addition, there are sometimes also radio/X-ray flares that show no obvious correlation. In this paper, we present simultaneous radio and X-ray observations of a promising tMSP candidate system, 3FGL~J1544.6$-$1125, which shows optical, $\gamma$-ray, and X-ray phenomena similar to PSR~J1023+0038, but which is challenging to study because of its greater distance. Using simultaneous VLA and {\it Chandra} observations we find that the radio and X-ray emission are consistent with being anti-correlated in a manner similar to PSR~J1023+0038. We discuss how our results help in understanding the origin of bright radio emission from tMSPs. The greater sensitivity of upcoming telescopes like the Square Kilometre Array will be crucial for studying the correlated radio/X-ray phenomena of tMSP systems.

\end{abstract}


\section{Introduction}\label{sec:into}
%%%%%%%%%%%%%%%%%%%%%%%%%%%%%%%%%%%%%%%%%%%%%%%%%%%%%%%%%%%%

%WHy it is important to study MSPs and where are they come from - tMSPs
Fast-spinning, rotation-powered millisecond pulsars (RMSPs) are valuable laboratories for fundamental physics experiments (e.g., gravity tests \citealt{Kramer2021,Archibald2018}; constraining the equation-of-state of dense matter \citealt{Demorest2010}; and low-frequency gravitational wave studies \citealt{Nanograv2023}). Understanding the origin of RMSPs is, thus, of great importance as it can provide insights into the minimal rotational period they can attain, their maximum mass, and the evolution of their magnetic field. 

The prevalent theory for the origin of RMSPs is the pulsar recycling mechanism \citep{Alpar1982,Radhakrishnan1982}. In this scenario, an old (spun-down) pulsar can be rejuvenated and spun-up to millisecond rotational periods through the transfer of angular momentum and mass via disk-fed accretion from the low-mass binary companion star --- during the low mass X-ray binary (LMXB) phase of its evolution. This scenario was supported by the discovery of accreting millisecond X-ray pulsars (AMXPs; see \citealt{Wijnands1998}) --- systems that exhibit coherent X-ray pulsations, which illustrated that material from the accretion disk was being channelled directly onto the magnetic polar caps of the neutron star). Moreover, in the last decade three systems --- PSR~J1023$+$0038, hereafter J1023 \citep{archibald2009, Stappers2014}; PSR~J1824$-$2452I (also known as M28I; \citealt{Papitto2013}), and PSR~J1227$-$4853 (also known as XSS~J12270$-$4859; \citealt{Bassa2014}) --- have been observed to transition between accretion-disk-dominated LMXB-like and disk-free RMSP states on timescales of months to years, directly confirming the evolutionary link between LMXBs and RMSPs. These systems are thus called transitional millisecond pulsars (tMSPs). Since 2013, XSS~J12270$-$4859 and M28I have remained in the RMSP state, while J1023 has persisted in the LMXB-like state.

%LMXB state - not like other LMXBs; X-ray moding
The LMXB-like state of the three known tMSPs is quite different from that of typical Galactic LMXBs with black hole (BH) or neutron star (NS) primaries. Firstly, tMSPs are the only X-ray binaries known to be bright $\gamma$-ray emitters. 
Secondly, two tMSPs --- J1023 and XSS~J12270$-$4859 --- have been observed to stay in the intermediate X-ray luminosity regime with \lx$\sim10^{32-34}$\,erg~s$^{-1}$ for years (similar \lx\ to that of very faint X-ray transients; \citealt{Wijnands2006}), while the majority of known BH and NS-LMXBs reside in either of two states: (i) an active accretion state, called an X-ray outburst (with X-ray luminosities of \lx$ > 10^{36}$\,erg~s$^{-1}$) or (ii) a little-to-no accretion state, called X-ray quiescence (\lx$<10^{32}$\,erg~s$^{-1}$). Only one out of three confirmed tMSPs, M28I, has ever been observed to reach a luminosity typical of outbursting LMXBs \citep{Papitto2013}. 

In their intermediate X-ray luminosity state (often called the sub-luminous X-ray state), the X-ray emission of J1023 \citep{Archibald2015} and XSS~J12270$-$4859 \citep{deMarino2013} have shown peculiar moding behaviour\footnote{It is worth noting that the X-ray lightcurve of M28I obtained at similar intermediate luminosities is also bimodal \citep{Linares2014}, but the duty cycles of the high and low modes are much longer (several hours).}: they exhibit a bi-stable X-ray light curve (see Figure~\ref{fig:J1023_compare}) characterised by rapid switching between X-ray `low' (\lx$ \sim 5 \times 10^{32}$\,erg~s$^{-1}$) and `high' modes (\lx$ \sim 5 \times 10^{34}$\,erg~s$^{-1}$) on minutes timescales. Occasional flares are also seen, with X-ray luminosities reaching up to \lx$ \sim 3 \times 10^{34}$\,erg~s$^{-1}$. This moding is also seen in optical \citep{Shahbaz2015} and UV \citep{Jaodand_uvj1023, Miraval-Zanon2022} lightcurves of tMSPs and is a unique feature of tMSPs among NS- and BH-LMXBs, although a similar moding behaviour was recently observed in an accreting white dwarf system with longer timescales (from 30 minutes to hours; \citealt{Scaringi2022}).

%X-ray pulsations
All three tMSP systems show coherent X-ray pulsations in their LMXB-like state. In the case of J1023 and XSS~J12270$-$4859 these pulsations (at the 1.69-ms spin period in both systems) have been observed at much lower X-ray luminosities (\lx $\sim$10$^{33}$\,erg~s$^{-1}$; \citealt{Archibald2015}) than previously observed in other AMXPs. In both J1023 and XSS~J12270$-$4859 coherent X-ray pulsations were only detected in the high X-ray mode and appear to switch off during the low X-ray mode. In J1023 they were also detected in the UV \citep{Jaodand_uvj1023, Miraval-Zanon2022} and optical \citep{Ambrosino2017} bands, again only in the high X-ray mode,\footnote{Optical pulsation were also detected in \jttt\ with reduced flux during X-ray flares \citep{Papitto2019}.} and appear to be exceptionally stable (the same pulse profile persisting for years). Long-term X-ray timing of J1023 \citep{Jaodand2016, Burtovoi2020} showed that its rotation period is slowing down in its LMXB-like state and its spin-down is actually enhanced (by 4\%, compared to the RMSP state) possibly because of negative torques on the neutron star from accretion material being `propeller-ed' away and because the radio pulsar mechanism is still active despite the presence of the accretion disk \citep{CotiZelati2014}.

\subsection{Radio and X-ray observations of LMXBs}

It is now clear that the behaviour of the accreting matter in tMSPs during their persistent intermediate X-ray luminosity state differs significantly from what is seen in other types of LMXBs. The X-ray outbursts of Galactic LMXBs can be detected across all wavelengths, with high energies (optical, UV and X-rays) mostly (but not exclusively) originating from the accretion inflow (the accreting disk and corona) and low energy (mm and radio) originating from the accretion outflow (e.g., a collimated jet, wind or discrete outflow ejecta). Multi-wavelength observations are an effective way to trace the interaction between the in-falling and out-flowing accretion matter in LMXBs \citep{Corbel2003, FENBELGAL2004, MIGFEN2006, Russell2006}. 

%BH-LMXBs
The radio emission of BH-LMXBs behaves mostly in a predictable way. At lower luminosities, during quiescence or in the early part of outburst, BH-LMXBs are in a hard X-ray state (associated with the truncated accretion disk). During this state the radio emission is believed to originate from the optically thick collimated jet of accelerated out-flowing particles: it has a flat radio to mm spectrum ($\alpha\sim0$) where $S_{\nu} \propto \nu^{\alpha}$ and its radio luminosity scales as a power law with the X-ray luminosity (and thus mass accretion rate) of the binary \lr$\propto$\lx$^{0.59}$ \citep{Corbel2003,Gallo2018} over many orders of magnitude\footnote{This correlation holds for `standard track' BH-LMXBs such as V404~Cyg and GX~339$-$4, whereas the so-called `outlier' sources (such as H1743$-$322 \citealt{Coriat2011} and Swift~J1753.5$-$0127 \citealt{Plotkin2017}) follow a steeper correlation near the peak of their outbursts, but break to move horizontally back to the standard track near quiescence.} down to quiescent levels (see black dots in Figure~\ref{fig:lrlx}). In the soft X-ray state (associated with dominant X-ray emission coming from a bright accretion disk that extends close to the primary) the radio emission is usually undetected, as the radio jet appears to switch off and detach during the hard-to-soft state transition \citep{Russell2020}. Around this time, optically thin (steep spectrum) radio flares may be detected, which are associated with discrete jet ejections around the transition from the hard to soft state. 

%NS-LMXBs
In contrast, the radio emission behaviour of NS-LMXBs appears to be much more complex. Firstly, NS-LMXBs are on average about 20 times fainter in radio than their BH counterparts \citep[e.g.][]{MIGFEN2006,Gallo2018} and display a much broader scatter in radio luminosities \citep{Gusinskaia2020} at the same \lx. This difference cannot be explained by differences in the mass of the primary star alone. In the past decade, substantial effort was put into studying the radio emission from NS-LMXBs with the number of radio/X-ray measurements having increased by almost an order of magnitude \citep{Tudor2017, vandenEijnden2021, Panurach2021}, mostly thanks to the increased sensitivity of available radio interferometers, such as the Karl G. Jansky Very Large Array (VLA), Australia Telescope Compact Array (ATCA) and MeerKAT. Nonetheless, a substantial fraction of NS radio measurements are upper limits (more than 60\% for \lx$<10^{36}$\,erg~s$^{-1}$), which underlines that we do not yet have the capabilities to study the faintest NS jets. Moreover, it should be noted that most systems are only observed once or twice and such measurements provide limited to no insight into how the radio emission evolves over a large X-ray luminosity range.

From what we see now, there is no single model that can explain the production of radio emission in all classes of accreting neutron stars. The two classes that stand out the most are highly magnetised NSs and tMSPs, with the former being statistically fainter \citep{vandenEijnden2021} and the latter being statistically brighter \citep{Gallo2018} than the rest. The less-magnetised NS-LXMBs show radio/X-ray behaviour similar to that of BH-LMXBs but with fainter radio jets in the hard X-ray state and less pronounced jet quenching in the soft X-ray state \citep{Gusinskaia2017, vandenEijnden2021}. Given the observational limitations pointed out above, it is not clear whether there is a single \lr/\lx\ correlation for NS-LMXBs as found for BH-LMXBs. Nonetheless, using the available observational data, \citet{Gallo2018} proposed the shallower (compared to BHs) correlation \lr$\propto$\lx$^{0.44}$ for less-magnetised NS-LMXBs. 

%Limitations of NS-LMXB radio observations
The low luminosities of the radio counterparts of NS-LMXBs, together with the limited sensitivity of current instruments, prevent good measurements of the spectral and polarimetric properties of their radio emission. This complicates the study of the origin of the radio emission from NS-LXMBs (i.e., distinguishing between a collimated jet, a wind, or another type of outflow). Furthermore, and relevant for the present study, most previous \lr/\lx\ measurements are done in a quasi-simultaneous way (observations in the two bands may be up to three days apart). While this methodology can be justified for many LMXBs that do not seem to exhibit short-term variability, quasi-simultaneous observations are insufficient for studying more complex and rapid accretion inflow-outflow interactions (e.g., \citealt{Tetarenko2021}), and tMSPs are the most obvious example of this issue.

%Tab
%%%%%%%%%%%%%%%%
\begin{table*}
\caption[All observations]{VLA and {\it Chandra} observations of \fgl. All uncertainties are 1$\sigma$. }

\begin{minipage}{180mm}
{
\renewcommand{\arraystretch}{3}
\begin{tabular}{@{\extracolsep{25pt}}c|cccccc@{}}
\hline\hline
\multicolumn{6}{c}{Radio VLA (project code: 18A-398)} \\
\hline
Epoch & \makecell[c]{MJD}  & \makecell[c]{ Time\\ UTC } & \makecell[c]{$\nu$\\(GHz)} & \makecell[c]{X-ray\\ mode} & \makecell[c]{$S_{\nu}$ \\ ($\upmu$Jy)} & \makecell[c]{\lr$^b$ $\times 10^{27}$ \\(erg\,s$^{-1}$) \\ at 5 GHz }\\
\hline
  1 &\makecell[c]{58224} &  \makecell[c]{16 Apr 2018\\07:22:06 -- 12:59:08} & \makecell[c]{8-12} & \makecell[c]{all \\ low \\ high} & \makecell[c]{11.9$\pm$1.6\\14.2$\pm$2.9\\9.7$\pm$1.9} & \makecell[c]{0.8$\pm$0.1\\1.0$\pm$0.2\\0.7$\pm$0.2}\\
\arrayrulecolor{gray}\hline
 2 &\makecell[c]{58491} &   \makecell[c]{8 Jan 2019\\13:00:50 -- 18:48:18} &  \makecell[c]{4-8} & \makecell[c]{all \\ low \\ high} & \makecell[c]{28.6$\pm$2.1\\56.6$\pm$8.3\\16.1$\pm$7.6} & \makecell[c]{2.0$\pm$0.1\\4.0$\pm$0.6\\1.1$\pm$0.4}\\
\hline\hline
\multicolumn{6}{c}{{\it Chandra} X-ray ($0.5-8$ keV)} \\
\hline
Epoch & \makecell[c]{MJD} &  \makecell[c]{Time\\ UTC} & \makecell[c]{Obs. ID } & \makecell[c]{X-ray\\ mode} & \makecell[c]{Unabsorbed$^a$ \\ flux  $\times 10^{-12}$\\($\mathrm{erg\,s^{-1}\,cm^{-2}}$)}   &  \makecell[c]{\lx$^b$ $\times 10^{33}$\\(erg\,s$^{-1}$)}  \\
\hline
 1 & \makecell[c]{58224} & \makecell[c]{16 Apr 2018\\07:08:09 -- 12:59:39} & 20902 & \makecell[c]{all \\ low \\ high} & \makecell[c]{$2.9\pm0.1$\\ $<$1.8 \\ $3.0\pm0.1$ }  & \makecell[c]{$4.3\pm0.1$\\ $<$2.5 \\ $5.7\pm0.1$} \\
\hline
 2 & \makecell[c]{58491} & \makecell[c]{8 Jan 2019\\12:47:31 -- 18:50:01} & 20903 & \makecell[c]{all \\ low \\ high}  & \makecell[c]{$3\pm0.1$\\ $<1.2$ \\ $3.9\pm0.1$ }  & \makecell[c]{$4.3\pm0.2$\\ $<$1.7 \\ $5.5\pm0.1$} \\
\arrayrulecolor{black}\hline
\end{tabular}
}
\begin{flushleft}{
$^{a}$ Derived for 1--10 keV, using a simple absorbed power-law model with fixed $N_{\rm H} = 1.7 \times 10^{21}$\,cm$^{-2}$ and fitted $\Gamma=1.73\pm0.04$\\
$^b$ Assuming a distance of 3.45 kpc and a flat spectrum.\\
}\end{flushleft}
\end{minipage}
\label{tab:obs}
\end{table*}
%%%%%%%%%%%%%%%%%%%%%%%%%

\subsection{Multi-wavelength observations of confirmed tMSPs and tMSP candidates}\label{sec:into-rx_tmsps}

Studies of the radio emission of tMSPs in their LMXB-like state are scarce due to the lack of known and actively accreting tMSP systems.
Initially, tMSPs were targeted in much the same way as other LMXBs: via quasi-simultaneous X-ray and radio observations. This revealed that all three tMSPs are active radio sources in their accretion-disk-dominated LMXB-like state with radio emission being consistent with a compact, partially self-absorbed synchrotron jet (similar to what is often observed in other LMXBs). However, tMSPs tend to be much brighter in the radio than other NS-LMXBs, with radio luminosities comparable to those of BH-LMXBs \citep{DEL2015}. On closer inspection, the radio emission of J1023 in its sub-luminous X-ray state was observed to be rapidly variable (by a factor of a few on minutes timescales), while its flat radio spectrum was still consistent with an optically thick radio jet.

Currently, the only tMSP that is in an LMXB-like state is J1023. Thus, most of our knowledge about the complex behaviour of tMSPs comes from extensive follow-up studies of this single system. \citet{Bogdanov2018} performed a 5-hr-long, strictly simultaneous {\it Chandra} and VLA observation of J1023 that revealed that its variable radio emission also exhibits modes of high and low flux density and that they are strongly anti-correlated with the low modes of J1023's X-ray light curve. During each one of the nine low X-ray modes captured in that observational campaign, the radio flux density rapidly increased by a factor of a few at the start of the low X-ray mode and returned to its original level at the end of the X-ray low mode. Additional radio flaring was also observed in the 5-hr radio lightcurve of J1023: one flare that seems to be related to an X-ray flare and multiple `orphan' radio flares that have no X-ray counterpart and occur during the X-ray high mode. More extensive multi-wavelength campaigns have been carried out to study this complex modality of J1023's emission (\citealt{Jaodand_uvj1023, Baglio2023}), revealing that J1023 is `moding' at nearly all wavelengths and X-ray modes are correlated all the way from hard X-rays to UV wavelengths and switch to being anti-correlated from mm waves to radio.

The simultaneous multi-wavelength observational campaigns on J1023 \citep{Bogdanov2018, Baglio2023} also uncovered intriguing spectral behaviour of its radio emission. During the radio mode flaring, the high radio frequencies clearly lead the low frequencies --- typical for an expanding, synchrotron-emitting plasma. A joint analysis of all the X-ray low modes shows that the average spectral index evolves from optically thick to thin from the first to the second half of an X-ray low mode ($\alpha_{1} =0.67 \pm 0.12$ to $\alpha_{2}=-0.27 \pm 0.05$), indicating an evolving synchrotron spectrum. More puzzling is the fact that the duration of the X-ray low mode matches that of the radio flare. This is most straightforwardly explained if there is a causal connection between the ejected plasma and events at the inner edge of the accretion disk (and the magnetosphere of the NS). The classic collimated jet-like outflow (a common interpretation of radio emission from NS-LMXBs) would not have a sufficiently fast reaction time to match the X-ray mode switches so closely. 

\citet{Baglio2023} suggest that the radio emission from \jttt\ originates from two separate mechanisms: (i) an optically thick jet that is active at all times and is responsible for the radio emission visible during the high X-ray modes, and (ii) discrete mass ejections in the low X-ray modes that are caused by rapid stripping and expulsion of the inner-most material of the accretion disk. Those ejections, however, cannot explain the `orphan' radio flares that occur during the high X-ray modes. An alternative scenario, proposed by \citet{Bogdanov2018}, involves an active radio pulsar: the radio flares are attributed to the expulsion of plasma from the pulsar's magnetosphere or the inflation of a short-lived compact pulsar wind nebula. In both cases, it is still unknown what can cause such rapid changes in or near the magnetosphere of the NS. 

Understanding the mechanisms behind the unusual outflow behaviour of tMSPs can bring valuable insights into how the NS magnetosphere interacts with the inflowing accreting matter, providing more context for the pulsar recycling scenario \citep{Alpar:1982}. However, most of what we know about the radio emission of tMSPs comes from only one system (J1023), and the question remains whether this behaviour is common to all tMSPs or the peculiarities of this individual system. Multiple properties of tMSPs that happen to be unique to this class (X-ray moding, $\gamma$-ray counterpart, etc.) make it easier to identify new systems that are potential tMSP candidates \citep{Jaodand2018}. The most straightforward way of identifying such candidates is the cross-association of {\it Fermi} $\gamma$-ray sources with known X-ray sources. A handful of sources have been found in this way and most of them indeed appear to be rapidly variable X-ray sources with similar intermediate X-ray luminosities \citep{Papitto2022}. 

A few strong X-ray/$\gamma$-ray candidates have been followed up with multi-wavelength X-ray and radio campaigns to investigate their accretion inflow-outflow interactions, all displaying a complex, variable behaviour in both radio and X-ray. CXOU~J110926.4$-$650224, as reported by \citet{CotiZelati2019}, shows clear X-ray moding light curves with high, low and flare X-ray modes. The radio counterpart of this source is undetected at high radio frequencies (5.5 and 9\,GHz) with ATCA \citep{CotiZelati2019, CotiZelati2021}, but MeerKAT observations at low frequencies ($\sim1.28$\,GHz) revealed its bright radio counterpart (33\,$\mu$Jy on average, with up to 266\,$\mu$Jy flares; \citealt{CotiZelati2021}), with radio luminosities comparable to those of BH-LMXBs and consistent with those of other tMSPs (see Figure~\ref{fig:lrlx}). The observed variable radio flux density of CXOU~J110926.4$-$650224 was not found to be in anti-correlation with the X-ray low modes\footnote{This non-detection is not due to sensitivity constraints. An anti-correlated variability similar to that of \jttt\ would have been detected with $\gtrsim 10 \sigma$ significance \citep{CotiZelati2021}.}, but instead potentially linked to the X-ray flaring in the system \citep{CotiZelati2021}. Two other candidates 3FGL~J0427.8$-$6704 \citep{Kennedy2020} and NGC~6652B \citep{Paduano2021} were also observed to be rapidly variable in X-rays, but instead of showing clear low and high X-ray moding, they tend to stay in the flare-dominated mode. They are both radio-loud sources (4-8 GHz flux densities of $\sim$0.3 mJy and $\sim$0.07 mJy for 3FGL~J0427.8$-$6704 and NGC~6652B, respectively) that also populate the same space as BH-LMXBs in the \lr/\lx\ diagram. The radio emission of 3FGL~J0427.8$-$6704 \citep{Li2020}, similarly to that of CXOU~J110926.4$-$650224, seems to brighten during X-ray flares, showing no anti-correlation with the X-ray flux\footnote{As with CXOU~J110926.4$-$650224 \citep{CotiZelati2021}, this non-detection is not limited by sensitivity, as the source has a relatively high radio flux density; see Table 3 of \citet{Li2020}.}. NGC~6652B, observed by \citet{Paduano2021}, showed hints of anti-correlation between radio and X-ray fluxes: in this case the radio flux density seems to be lower during X-ray flares of the source compared to its level in between the flares. Both the radio and the X-ray luminosity of NGC~6652B are much brighter than in other tMSPs in the low-luminosity state (see Figure~\ref{fig:lrlx}). Finally, a multi-wavelength study of the very strong tMSP candidate 3FGL~J1544.6$-$1125 is the subject of this paper.

%Fig
%%%%%%%%%%%%%%%%%%%%%%%%%

\begin{figure*}
\centering
\includegraphics[width=0.97\textwidth]{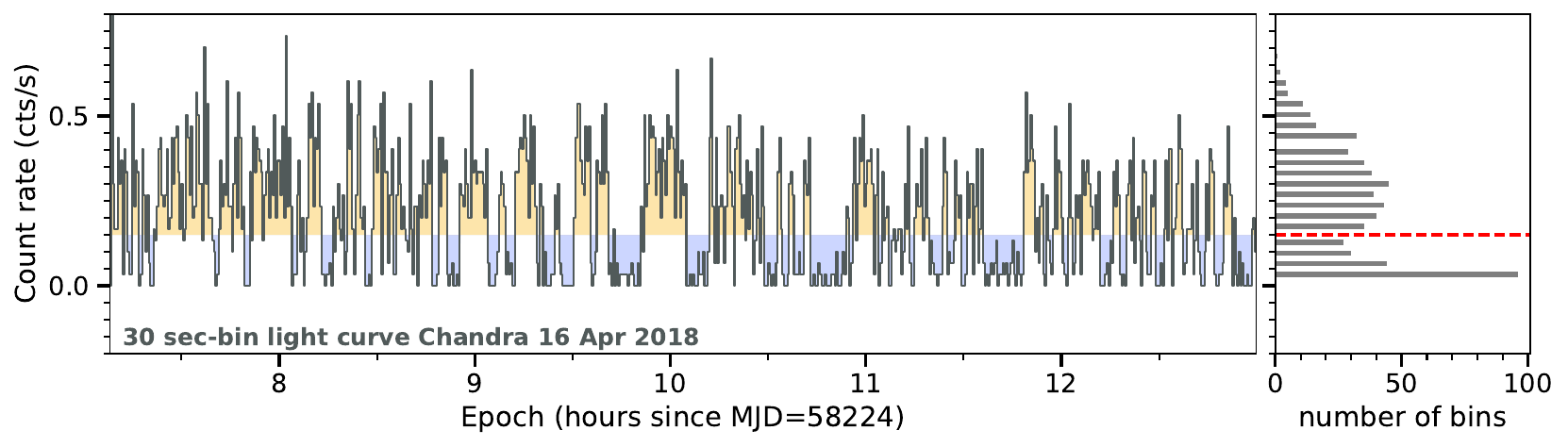}
\includegraphics[width=0.97\textwidth]{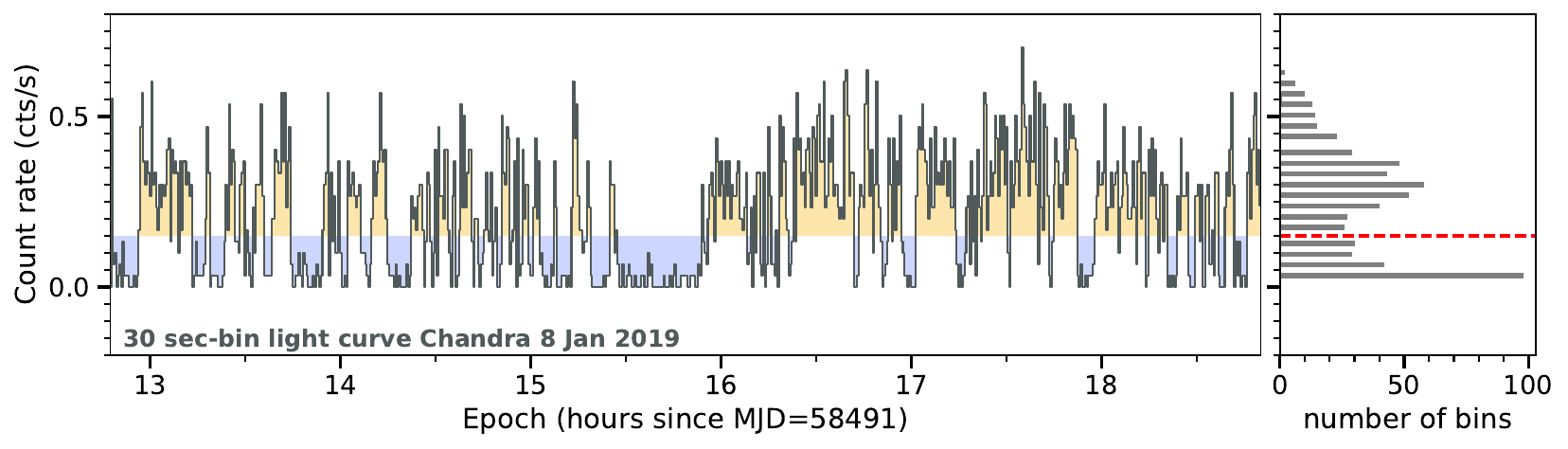}

\caption{{\bf Left panels:} {\it Chandra} $0.5 - 8$\,keV (grey) background-subtracted count rate time series of \fgl\ grouped in 30-s intervals from the Epoch~1 ({\it top}) and Epoch~2 ({\it bottom}). The data show abrupt changes of the X-ray flux with clear low modes.  Highlighted in blue and yellow are parts of the X-ray lightcurves for when we classify the source in a low or high X-ray mode, respectively. No obvious X-ray flares (typical for known tMSP) are present.
 {\bf Right panels:} Distribution of X-ray count rate per interval bin. The horizontal red dashed line corresponds to the count rate separation used to distinguish between the low and high mode in our analysis.}\label{fig:xray_lcs}
 
\end{figure*}
%%%%%%%%%%%%%%%%%%%%%%%%% 

\subsection{3FGL~J1544.6$-$1125}\label{sec:into-j1544}

3FGL~J1544.6$-$1125 (hereafter \fgl; found by \citealt{BogHal2015}) is one of the most promising tMSP candidate systems. It was discovered through a targeted search of previously unassociated {\it Fermi} sources. 
\fgl\ was found to be a faint X-ray source (flux $\sim 10^{-12}$ erg s$^{-1}$ cm$^{-2}$), with an X-ray spectrum well described by an absorbed power-law with a photon index of $\Gamma\approx$1.7, similar to other tMSPs.
The optical spectrum of \fgl\ displayed notable H and He emission lines, indicating the presence of an accretion disk. Archival X-ray and radio data showed that \fgl\ had not transitioned to an RMSP state over the course of at least 15 years prior to its discovery \citep{Bogdanov2016}. The {\it XMM-Newton} X-ray light curve of \fgl\ demonstrated distinct high and low modes over a timescale of tens of seconds \citep{Bogdanov2016}. The moding behaviour was also observed in the optical lightcurve with $\sim0.5$\,mag amplitude variability together with sporadic optical flaring of $\sim1-1.5$\,mag. Optical time-resolved spectroscopy revealed a period of 5.8\,hr \citep{Britt2017}, indicating that \fgl\ is in a binary system with a main sequence companion of intermediate mass (0.2--0.7 $\msun$; similar to the 3 confirmed tMSPs) and a likely face-on orientation ($i\sim5-8^{\circ}$). Photometric modelling of \fgl's companion placed the system at a distance of 3.8$\pm$0.7\,kpc \citep{Britt2017}, which is consistent with the recent distance estimate from the {\it Gaia} DR3 parallax of 3.1$\pm$0.5\,kpc \citep{GaiaDR3, BailerJones2021}. Finally, quasi-simultaneous VLA and {\it Swift}-XRT observations \citep{Jaodand2021} revealed radio emission from the source at levels of up to 40\,$\mu$Jy (in the 8--12 GHz band) that placed \fgl\ on the \lr/\lx\ diagram close to J1023 in the low-luminosity state. Importantly, this analysis showed that the radio flux density suddenly increases by a factor of $\sim2-3$ between hour-long observations separated by a few days. The radio flux measurements across these observations varied from $<13$, $<16$ (3$\sigma$ upper limits), 24, to 40\,$\mu$Jy. This inter-epoch variability is consistent with that observed in J1023 and could similarly be linked to radio/X-ray moding or sporadic radio flaring in the system. However, this analysis used quasi-simultaneous radio/X-ray observations, whereas truly simultaneous radio/X-ray observations are critical to investigate this further.

In this paper we present strictly simultaneous VLA and {\it Chandra} observations of \fgl. We use these observations to investigate the nature of its radio emission and to compare it with J1023 and other tMSPs. We also place the observed radio emission in the context of other accreting systems.  In \S\ref{sec:obs} we present the observations and data analysis. In \S\ref{sec:results} we report our findings and discuss their interpretation in \S\ref{sec:discussion}.

\section{Observations and data analysis}
\label{sec:obs}

In J1023, the radio and X-ray brightness can be coupled on very short timescales. In order to investigate this relationship in the candidate tMSP system \fgl\, we coordinated strictly simultaneous VLA and {\it Chandra} observations. We performed two observations, each 5-hr long: on 16 April 2018 (MJD~58224) and 8 January 2019 (MJD~58491), with VLA and {\it Chandra} observing time almost entirely overlapping (see Table~\ref{tab:obs} for more details).

\subsection{X-ray data: {\it Chandra}}

\fgl\ was observed with {\it Chandra} on 16 April 2018 (ObsID 20902) and 8 January 2019 (ObsID 20903) for a duration of $\sim$21\,ks each. In both observations the ACIS-S spectrometer was used in a continuous-clocking read-out mode in order to obtain 1-D images and enable high time resolution\footnote{The fast readout was necessary to avoid event pileup that would occur in full imaging mode \url{https://cxc.cfa.harvard.edu/ciao/ahelp/acis_pileup.html}}. The data were analysed using the Chandra Interactive Analysis of Observations software, {\tt CIAO} v4.15 \citep{ciao2006}, and the accompanying calibration database, {\tt CALDB} 4.10.4. The {\tt chandra\_repro} pipeline was run on both observations to ensure the correct calibration. After restricting the CCD energy range to $0.5-8$\,keV (using the task {\tt dmcopy}), the source events were extracted from a rectangular region with 5$\arcsec$ width around the source position taken from \citet{GaiaDR3,Gaiamission}. The background events were extracted from a rectangular region with the same size located $\sim$10\arcsec\ away from the source position along the image line. The lightcurves were created using the {\tt dmextract} task with 30-s binning (this binning was chosen to balance signal-to-noise ratio and time resolution). For each epoch, we subtracted the background lightcurve from the source lightcurve to obtain the background-subtracted lightcurve of \fgl\, see Figure~\ref{fig:xray_lcs}. The event timestamps were translated from the Terrestrial Time (TT) standard used by {\it Chandra} to UTC to enable alignment with the radio light curve \footnote{{\it Chandra} timestamps were not barycentered, resulting in a 0.05--0.5s difference of \fgl's signal arrival times at {\it Chandra} and VLA, which is significantly shorter than the time bins of our lightcurves.}. 

%Fig
%%%%%%%%%%%%%%%%%%%%%%%%%
\begin{figure}
\centering
\includegraphics[width=0.46\textwidth]{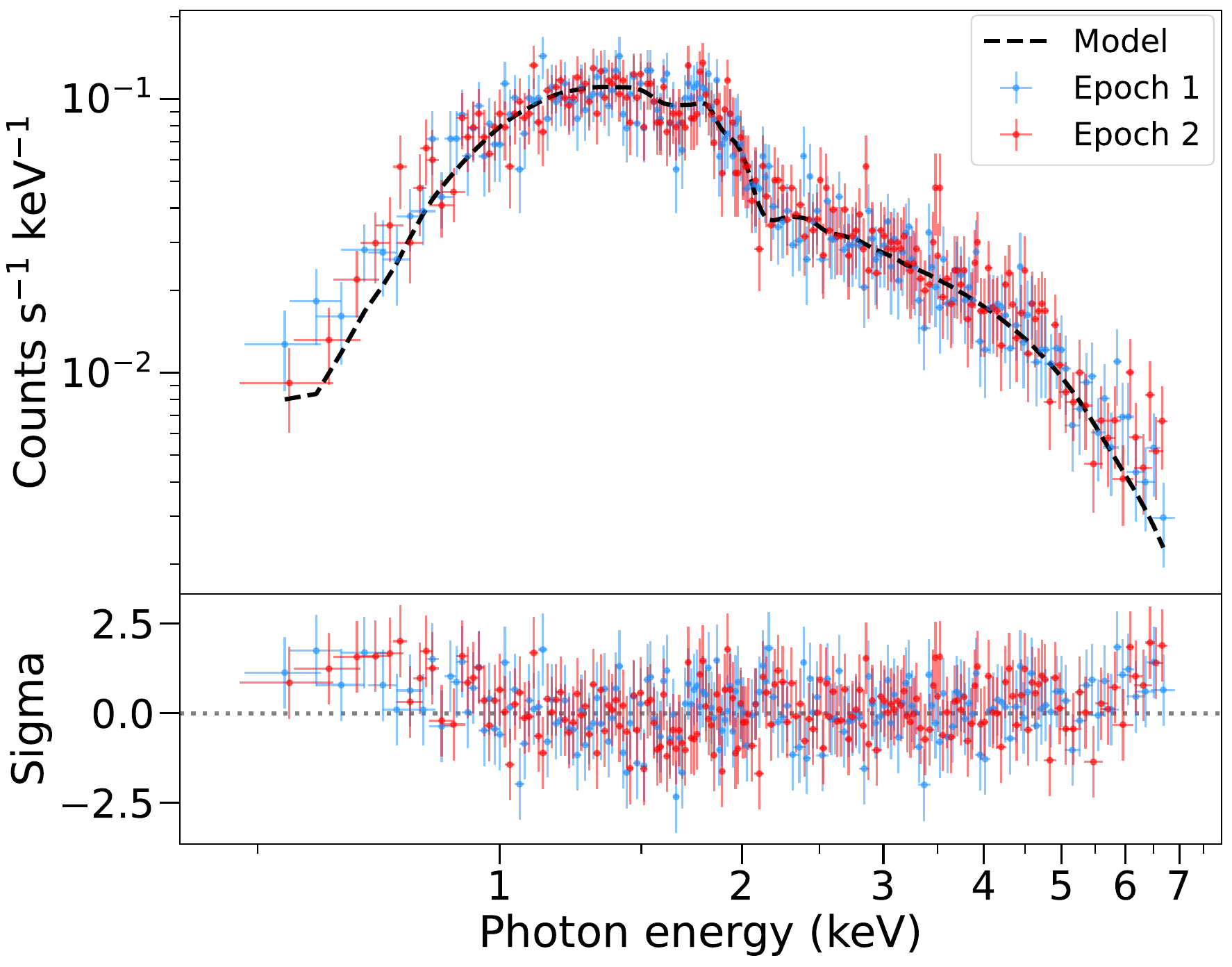}
\caption{{\it Top:} X-ray spectra of \fgl\ obtained from our two {\it Chandra} observations. Spectra is almost identical between two epochs: blue and red points correspond to the first and the second epoch respectively. Both spectra were fitted separately with the simple absorbed power-law model (shown with the black dashed line) with the neutral Hydrogen density, $N_{\rm H}=1.7 \times 10^{21} cm^{-2}$ , fixed to the value reported in \citet{Bogdanov2016}. {\it Bottom:} Residuals of the spectral model fitting for both spectra. We exclude fitting of the data below 1 keV given changes to the ACIS effective surface area. }
\label{fig:xrayspectrum}
\end{figure}
%%%%%%%%%%%%%%%%%%%%%%%%% 
\subsubsection{X-ray Spectral Analysis}
We produced X-ray spectra for both observations to identify any spectral changes between the two epochs as well as previous observations of \fgl. We extracted spectra with the {\tt specextract} task using the same 5$\arcsec$ source region as above, but a larger $\sim 20\arcsec$ region for the background. To perform the spectral fitting, the spectra were grouped to include at least 15 photons per bin restricted to $0.5-7$\,keV. We fit data using the {\tt sherpa} tool from {\tt CIAO}. Each spectrum was loaded together with ancillary and response files produced by the reprocessing pipeline, and was background-subtracted using the background spectrum file we generated. We used a simple {\tt XSpec} absorbed power-law emission model to perform the fit ({\tt TBabs*powerlaw}) for both observing epochs as it was found to be sufficient to model the observed spectrum (see Figure~\ref{fig:xrayspectrum}). First, we left the hydrogen column density parameter $N_{\rm H}$ free, which resulted in $N_{\rm H}<3 \times 10^{20} \mathrm{cm}^{-2}$, $\Gamma = 1.51 \pm 0.06$, $\chi^2_n = 0.7$ (188 dof) and $N_{\rm H}=6\pm 3 \times 10^{20} \mathrm{cm}^{-2}$, $\Gamma = 1.54 \pm 0.06$, $\chi^2_n = 0.6$ (178 dof) for the first and second epoch, respectively. Since {\it Chandra} CCDs (especially ACIS) have a molecular contamination issue as well as the PSF broadening effect up to $0.8-1$\,keV \footnote{\url{https://cxc.cfa.harvard.edu/proposer/POG/}} we decided to fix the $N_{\rm H} = 0.17 \times 10^{22} \mathrm{cm}^{-2}$, as previously measured for \fgl\, from {\it XMM-Newton} data \citep{Bogdanov2016}. These fits yielded $\Gamma = 1.72 \pm 0.04$, $\chi^2_n = 0.91$ (189 dof) and $\Gamma = 1.73 \pm 0.04$, $\chi^2_n = 1.0$ (179 dof) for the first and second epochs, respectively. These fixed-$N_{\rm H}$ models were then used to extract the unabsorbed $1-10$\,keV fluxes for both observations using the {\tt cflux} modification {\tt TBabs*cflux*(powerlaw)}. The obtained X-ray fluxes are listed in Table~\ref{tab:obs} under X-ray mode `all'. Separate high and low X-ray mode fluxes were also estimated by scaling the obtained mean fluxes according to the change of {\it Chandra}'s count rate in low and high modes. Since the count rate of \fgl\ is nominal in the low X-ray mode, we could only place an upper-limit on the low X-ray mode flux.

%Fig
%%%%%%%%%%%%%%%%%%%%%%%%%

\begin{figure}
\centering
\includegraphics[width=0.45\textwidth]{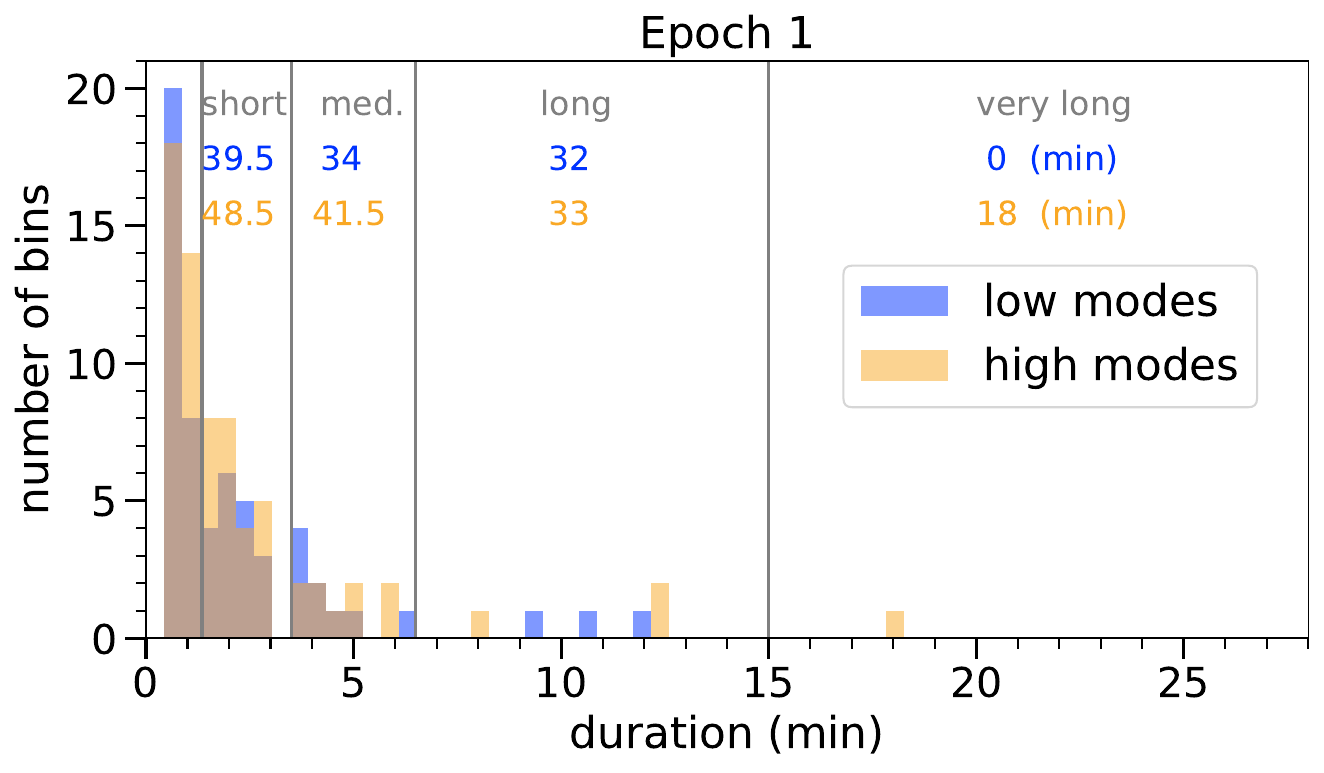}
\includegraphics[width=0.45\textwidth]{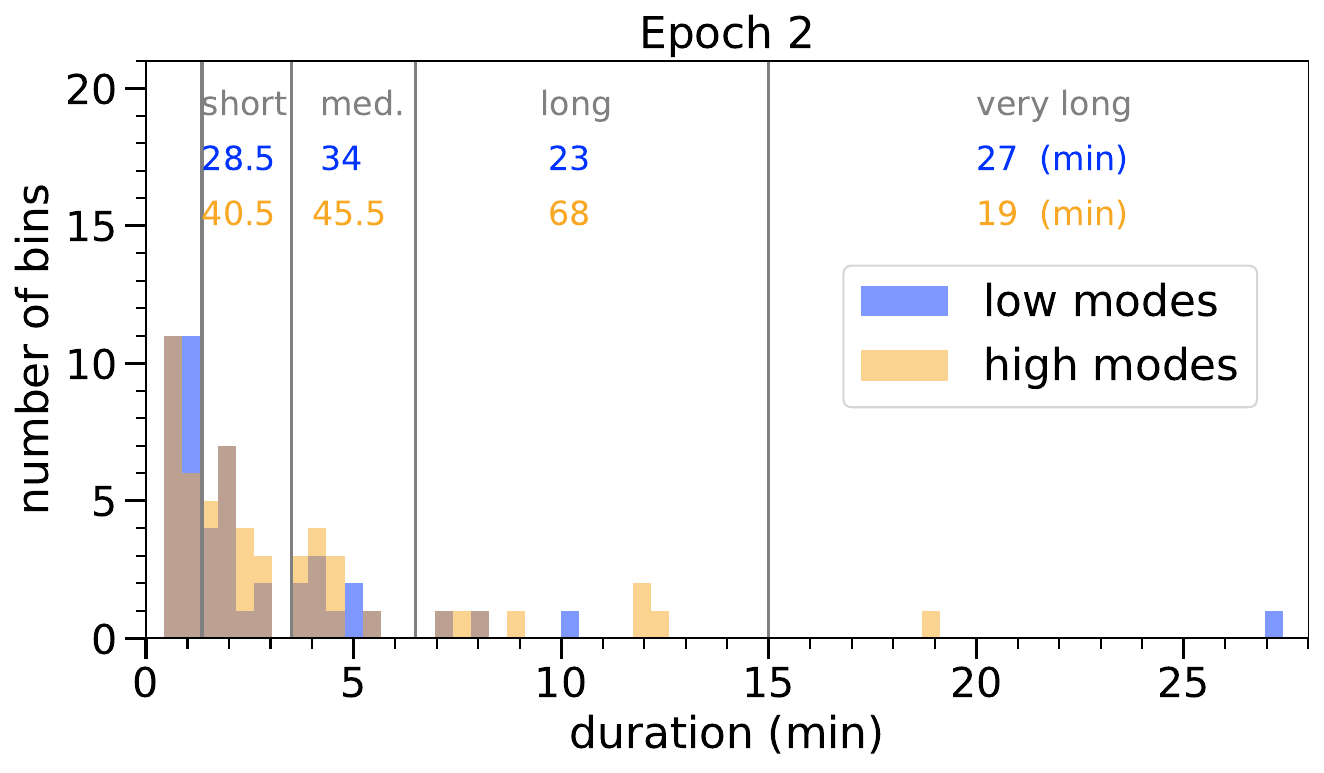}

\caption{Distribution of low and high mode duration in Epoch~1 ({\it Top}) and Epoch~2 ({\it Bottom}). During the first epoch \fgl\ had more rapid variability resulting in a larger number of short-duration modes than in the second epoch. The vertical lines correspond to chosen selections of mode duration intervals used to evaluate the dependence of radio flux density on X-ray mode duration (see Figure~\ref{fig:radio_modes_duraion} and Section~\ref{sec:res-radio}). The total time (in minutes) summed over all modes in each interval is listed at the top of the plot with blue and orange numbers corresponding to low and high modes respectively.}
\label{fig:xraymodes}
\end{figure}
%%%%%%%%%%%%%%%%%%%%%%%%% 

\subsection{Radio data: VLA}

%Fig
%%%%%%%%%%%%%%%%%%%%%%%%%

\begin{figure*}
\centering
\includegraphics[width=0.9\textwidth]{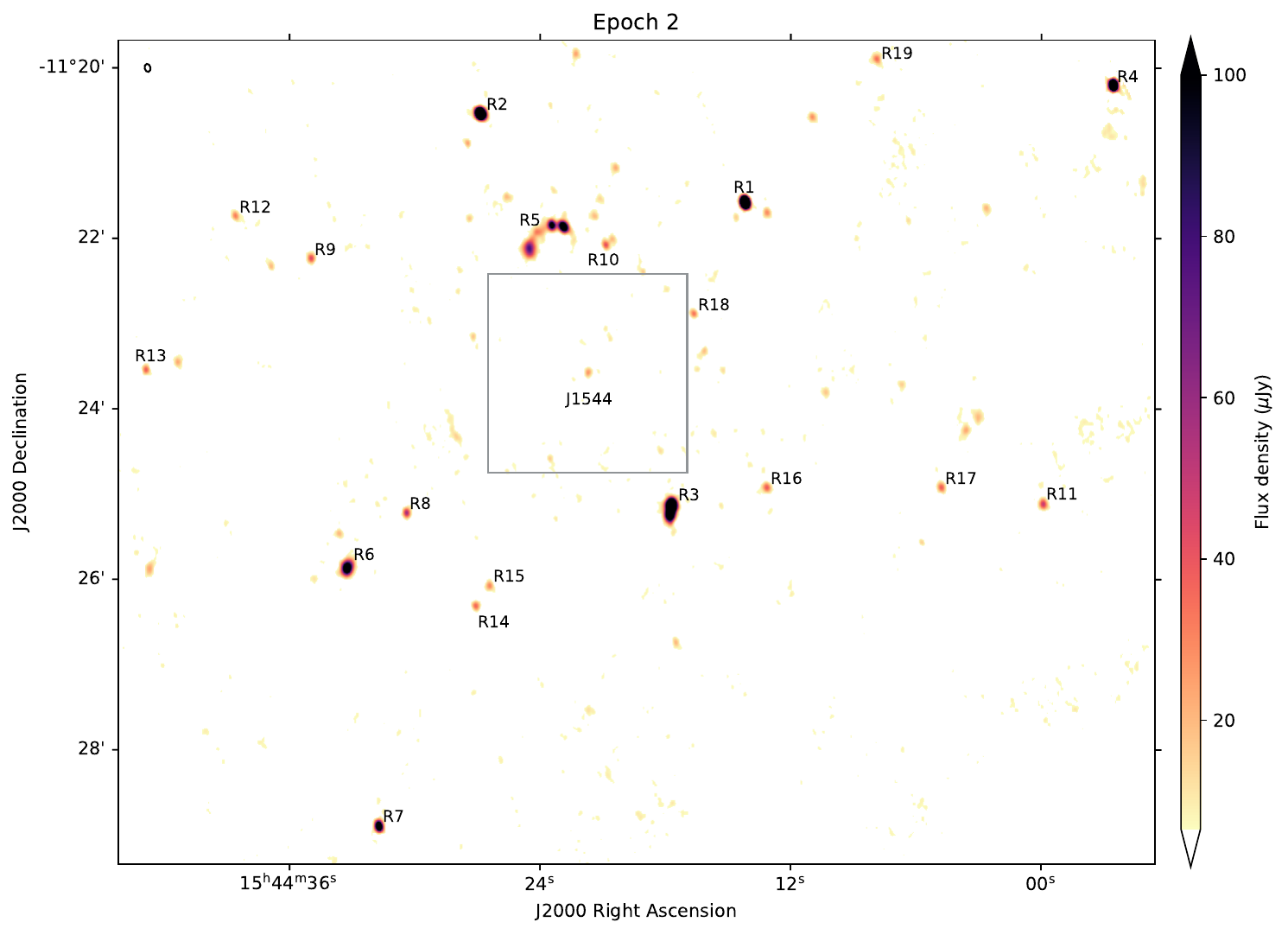}
\includegraphics[width=0.9\textwidth]{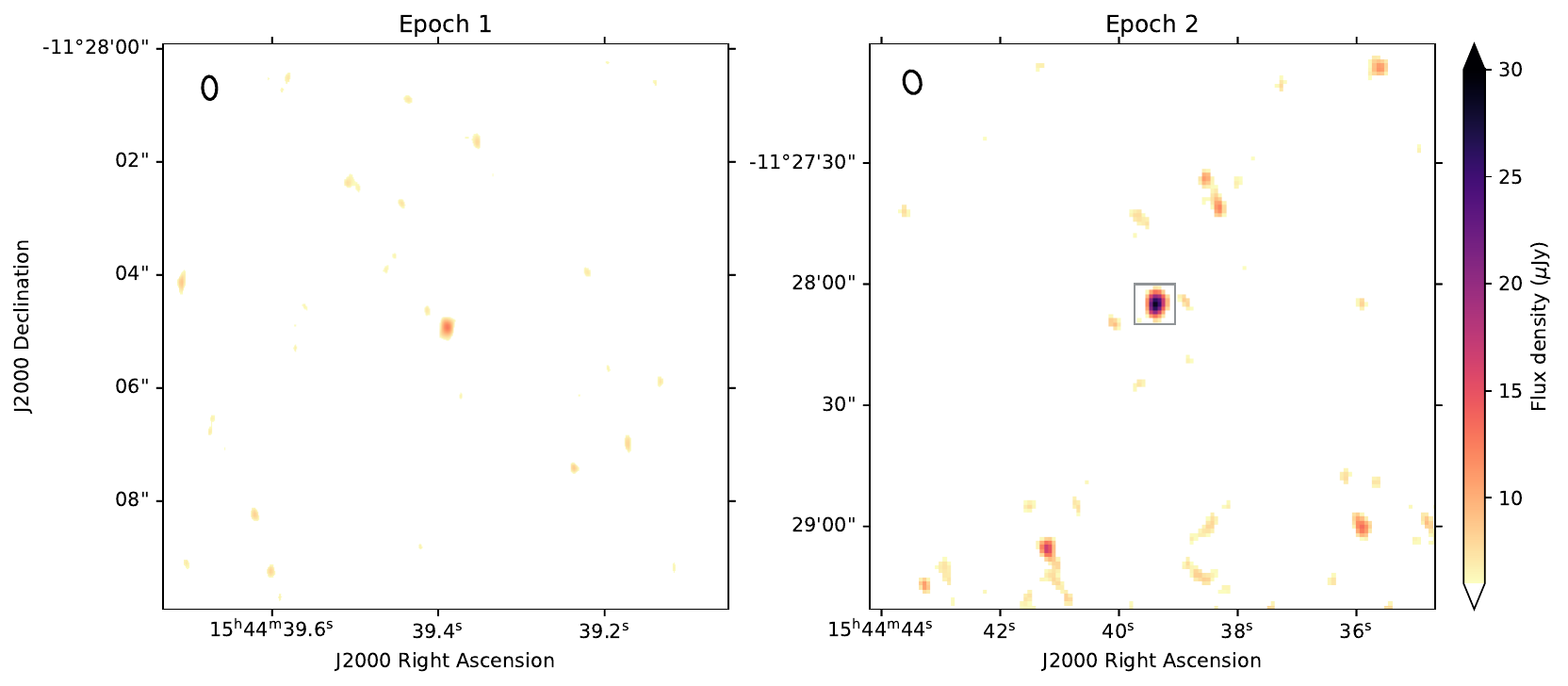}

\caption{Radio interferometric images (total intensity, Stokes~I) of both VLA observation epochs. {\it Top:} Widefield image of \fgl\ obtained by averaging all 5\,hr of Epoch~2. The grey square indicates the size of the zoomed-in region around the target shown in the bottom right panel. Marked background sources R1$-$R19 are used in a comparative analysis (see Figure~\ref{fig:variability} of the Appendix). Their average fluxes are ranging from 10 to 800 $\mu$Jy (the brightest sources R1-R7 are saturated in the image). {\it Bottom:} zoomed-in radio images of \fgl\ of the first ({\it left}) and second ({\it right}) radio observations (averaged over the full exposure time on source). Black ellipses in the top left corners corresponds to the average beam size and orientation at the time of the observation. The colour bar is common to the bottom two panels, with a range smaller than that of the colour bar in the top panel. Note the resolution difference between the two epochs as the first observation was performed in VLA A-configuration (synthesized beam size $0.41 \times 0.25$ \arcsec), with the second one in C-configuration (synthesized beam size $5.55 \times 4.15$ \arcsec). The grey square in the zoomed-in image of the second epoch indicates the size of the first epoch image shown on the bottom left panel. 
}\label{fig:radio_images}
\end{figure*}
%%%%%%%%%%%%%%%%%%%%%%%%% 

%Fig
%%%%%%%%%%%%%%%%%%%%%%%%%
\begin{figure*}
\centering
\includegraphics[width=\textwidth]{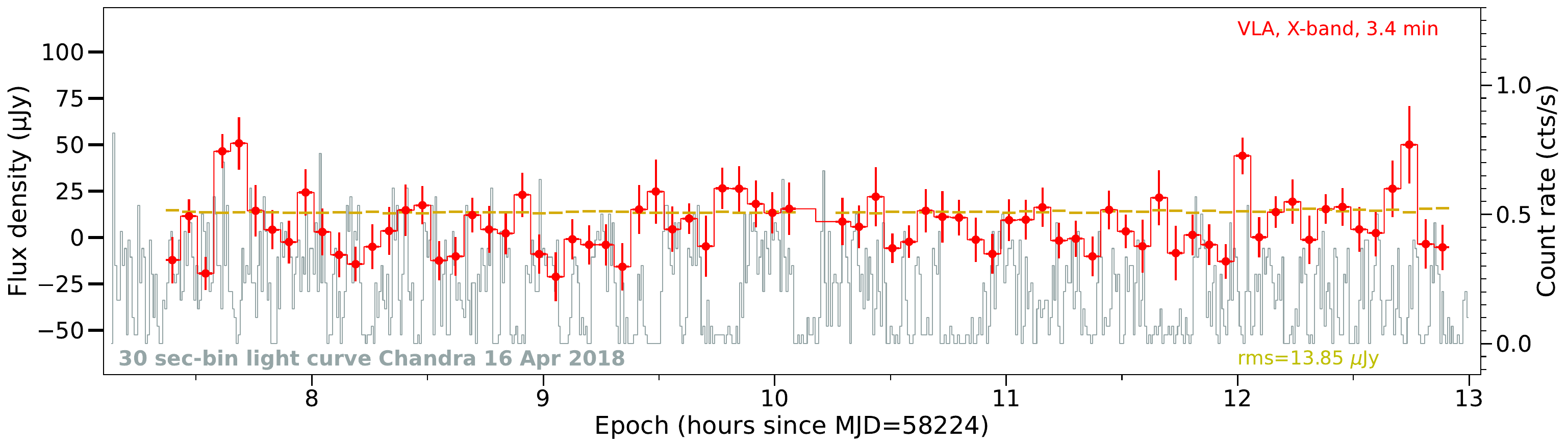}
\includegraphics[width=\textwidth]{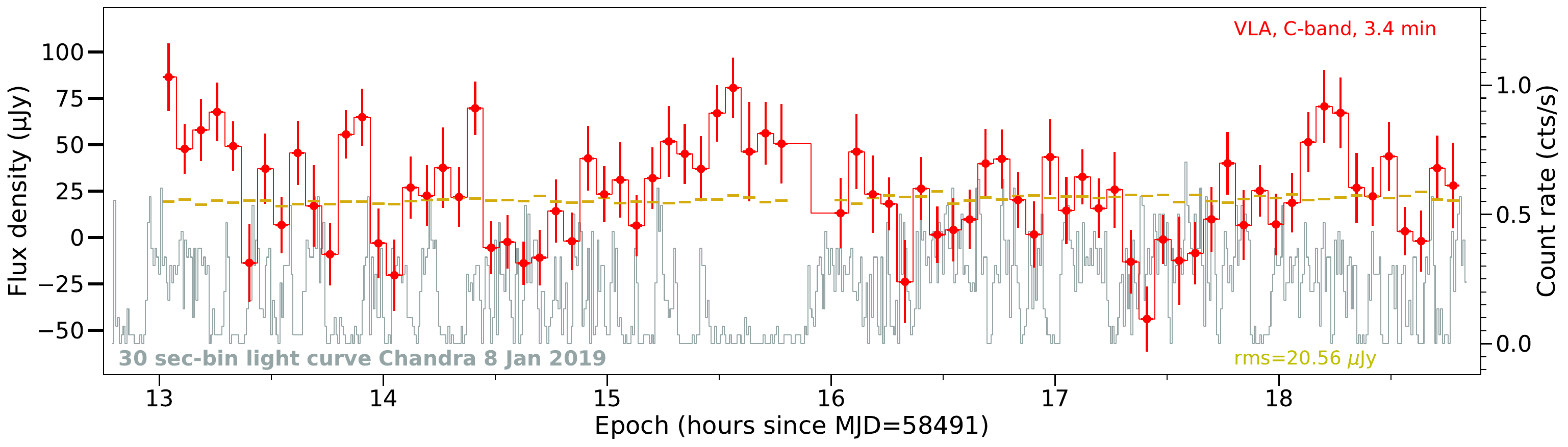}
\caption{Radio (VLA) and X-ray ({\it Chandra} $0.5-8$\,keV) lightcurves from Epoch~1 (\textit{top}) and Epoch~2 (\textit{bottom}). Radio lightcurves (binned at 3.4\,min) are shown with red symbols and their error-bars indicate the 1-$\sigma$ uncertainty on the flux density fit at the known source position (negative flux density values are consistent with the rms noise of the image and arise as part of the cleaning process). Horizontal yellow lines correspond to the off-source RMS noise of each individual image. The X-ray lightcuves (binned at 30 s) from both epochs (shown in Figure~\ref{fig:xray_lcs}) are over-plotted for each epoch with grey lines. During this first epoch the radio emission of \fgl\ appears to be constant and mostly consistent with the RMS noise for such short time-bins, except a few points at the beginning. During the second epoch \fgl\ had more visible variability in the radio, modulating on a time scale of $\sim10-30$ min.  Also the X-ray flux in the second epoch was more stable on short time-scales, staying in the same level for statistically longer times than in the first epoch. In particular, there is a very long X-ray low mode right in the middle of the observation. During this long low mode, \fgl\ was visibly brighter in radio. Furthermore, a hint of anti-correlation between X-ray and radio lightcurves is visible in the bottom panel: the source being brighter in radio when it is dimmer
in X-rays and vice-versa. There no obvious trends of radio emission being brighter in low-X-ray modes than that in high X-ray modes during the first epoch (top panel). \label{fig:radio_xray_lcs} }
\end{figure*}
%%%%%%%%%%%%%%%%%%%%%%%%%

Two 5-hr observations with the VLA were performed at the same time as (and entirely overlapping with) the {\it Chandra} observations, on 16 April 2018 and 8 January 2019 (project code: 18A-398). During the first epoch, the array was in A-configuration and the observation was taken in the X-band ($8-12$\,GHz). During the second epoch the array was in C-configuration and the observation was taken in the C-band ($4-8$\,GHz)\footnote{\url{https://public.nrao.edu/vla-configurations/}}. The observational setup and calibration were identical to those described in \citet{Jaodand2021}. We used 3C286 as the primary flux density calibrator (observed in the middle of the observation), and J1543$-$0757 as the phase calibrator with a cycle time of $\sim$4\,min. In the first epoch, \fgl\ was observed for a total time of 260\,min, via 75 $\sim$3.5-min scans. In the second epoch, \fgl\ was observed for a total time of 252\,min via 78 $\sim$3.4-min scans.

The VLA pipeline in CASA version 5.4.0 \citep{CASA2022} was used for initial calibration, followed by a small amount of additional manual flagging on the target field. Imaging was performed using the {\tt tclean} function in CASA 6.1.0. For the first epoch when the array was in A-configuration, the cleaning was done using outlier-fields: imaging a small region around the source (around 1~arcmin) and cleaning small patches around more spatially distant sources (see Figure~\ref{fig:radio_images}). During the second epoch the VLA was in C-configuration, meaning that at a pixel size of 1 \arcsec\ a single image of size $1500\times1500$ pixels was sufficient to image the full VLA primary beam.

First, for both epochs (separately), we imaged the entire observation (full time and frequency coverage of the epoch) to make the best mask of the surrounding field and determine the average radio flux density of \fgl. We used natural weighing and 6000 clean iterations, which were enough to achieve the expected RMS noise of $\sim2$\,$\mu$Jy/beam in the first epoch and $\sim3$\,$\mu$Jy/beam in the second epoch. We then used the produced mask to make time-binned images with the binning corresponding to durations of 1, 2 and 3 radio scans (3.5, 7 and 10.5\,min, respectively). Finally, we produced a number of images from specific time windows corresponding to different X-ray modes of \fgl\ during our observations. We made a set of images for each of the low and high X-ray modes (longer than 3.5\,min), for the average of all low-modes and all high modes (separately), as well as multiple images averaging only low or high modes with similar durations (the time-intervals for those were chosen based on the duration distribution of X-ray modes, and are indicated by vertical lines in Figure~\ref{fig:xraymodes}; each interval contains around half an hour of data). The source flux density was extracted using the CASA task \texttt{imfit} with a starting model fixed at the known source position (provided by \citealt{Jaodand2021,GaiaDR3}), while the image noise was estimated using an emission-free region of the image offset by several arcseconds from the source.

\section{Results}
\label{sec:results}
\subsection{X-ray}
\label{sec:res-xray}

\fgl\ is detected in the {\it Chandra} observations with total photon counts of 142 and 150 in the first and second epoch, respectively. In both the epochs the total number of counts from the background region are $\sim12$\,cts. The X-ray light curves show clear transitions between the low and high X-ray modes on a timescale of a few minutes (see Figure~\ref{fig:xray_lcs}). The distribution of count rate per time bin in both epochs (right panels of Figure~\ref{fig:xray_lcs}) appears bimodal for count rates greater than zero. A significant number of bins show count rates close to 0, from which we conclude that \fgl\ is too faint to be detected during the low X-ray mode. The count rate distribution appears bimodal with a local minimum around 0.16 cts/s (see red dashed line in the right panels of Figure~\ref{fig:xray_lcs}). Hence, we classified the X-ray lightcurve intervals to be in low or high mode based on the observed count rate staying lower or higher than this threshold for longer than 2 time bins.

We also find that the duration of high and low X-ray modes are variable in both epochs, ranging from short, 30-s modes (with duration limited by the time-bin resolution) to a long 28-min mode in the second epoch (Figure~\ref{fig:xraymodes}). In total, \fgl\ spent $T_{low}=$121\,min (34\% of the observation) and $T_{high}=$161\,min (45\% of the observation) during the first epoch, and $T_{low}=$118.5\,min (33\%) and $T_{high}=$183.5\,min (50\%) during the second epoch. These numbers are obtained including all duration modes, even the ones that are as short as 30\,s. If we exclude modes that are shorter than 3.5\,min (dictated by the time resolution of the radio light curves), the total time spent in each mode reduces to $T_{low} =66\, \mathrm{min}\, (19\%), T_{high}=92\, \mathrm{min}$\, (26\%) and $T_{low} =78\, \mathrm{min}\, (21\%), T_{high}=132\, \mathrm{min}$ (36\%) in the first and second epochs, respectively. The rest of the time is spent in the intermediate state rapidly switching between the modes.

During the first epoch, the X-ray mode changes are more frequent throughout the entire observation and \fgl\ spent more time in an intermediate state with a higher number of the shortest modes and a lower number of the longest modes than in the second epoch (see Figure~\ref{fig:xraymodes}). Nonetheless, the total time spent in low and high modes are very similar between the epochs, the former being nearly identical. No obvious X-ray flaring occurred in either of the two epochs.

We find that the X-ray spectrum of \fgl\ was also very similar during both epochs of our campaign, with a power-law index, $\Gamma$ of $\approx 1.7$, which is consistent with previous measurements from \citet{Bogdanov2016}. 

\subsection{Radio}
\label{sec:res-radio}

%Fig
%%%%%%%%%%%%%%%%%%%%%%%%%
\begin{figure}
\includegraphics[width=0.42\textwidth]{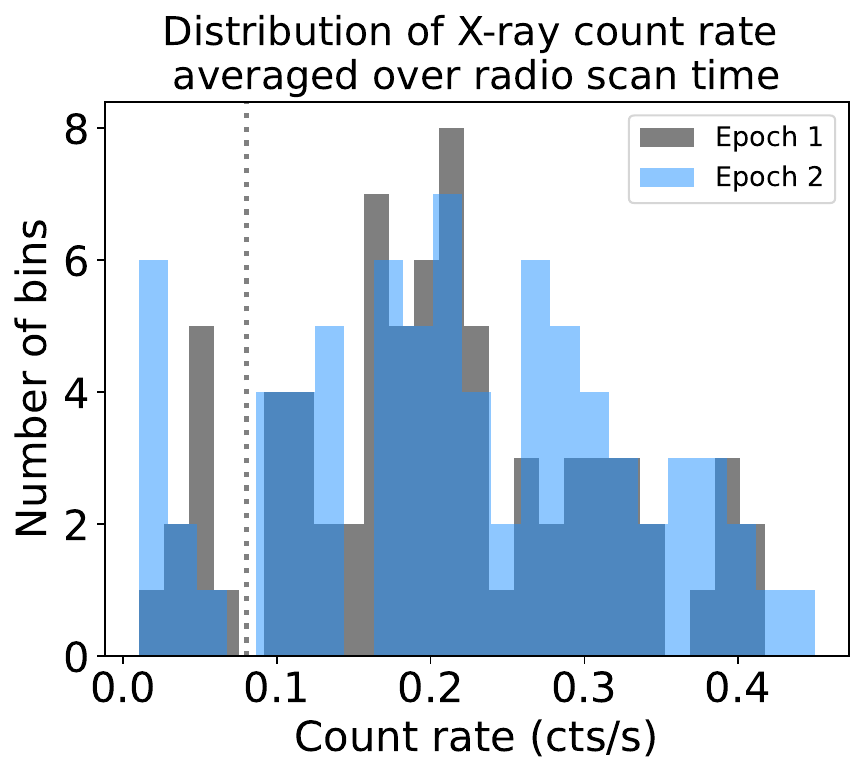}

\caption{Distribution of X-ray count rate averaged over radio scan time shown for both epochs. The vertical dotted grey line indicates the chosen separation associated with the X-ray low-mode lightcurve bins.}\label{fig:avx_cr_hist}
\end{figure}
%%%%%%%%%%%%%%%%%%%%%%%%%

%Fig
%%%%%%%%%%%%%%%%%%%%%%%%%
\begin{figure*}
\centering
\includegraphics[width=0.4\textwidth]{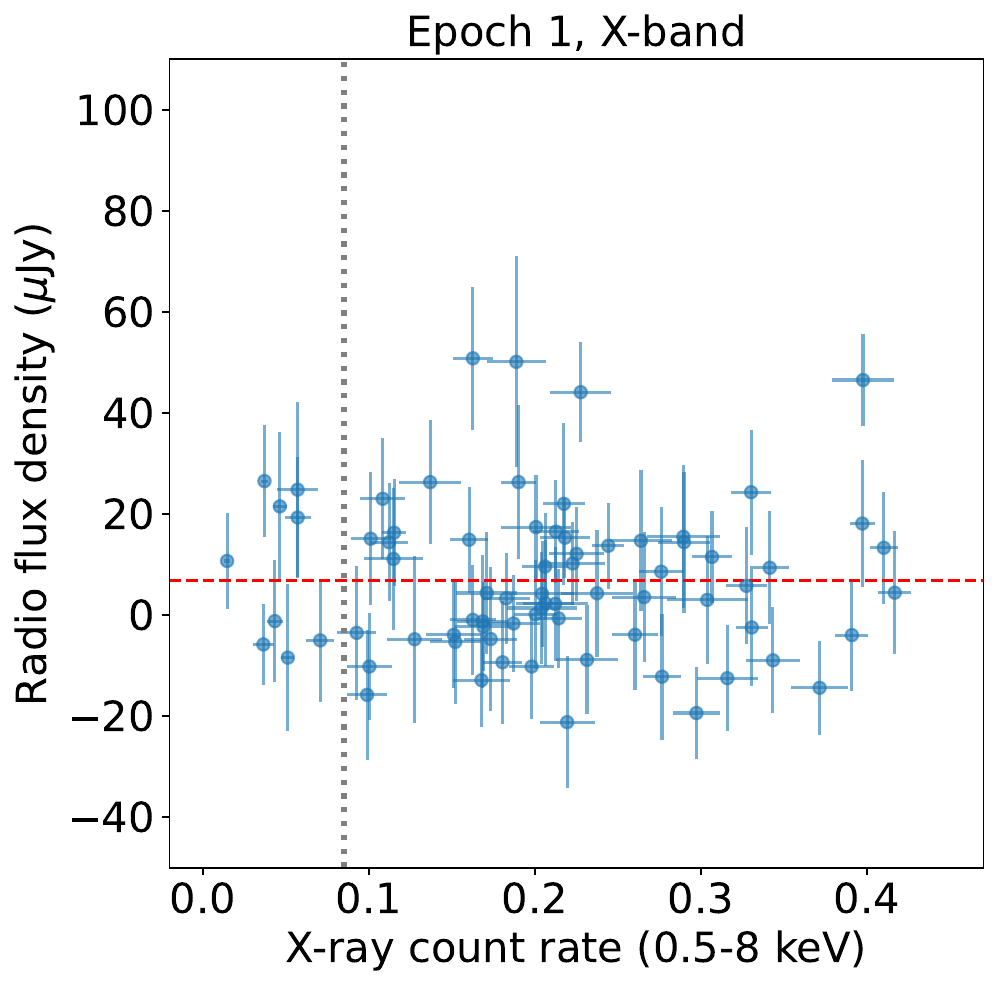}\hspace{5mm}
\includegraphics[width=0.4\textwidth]{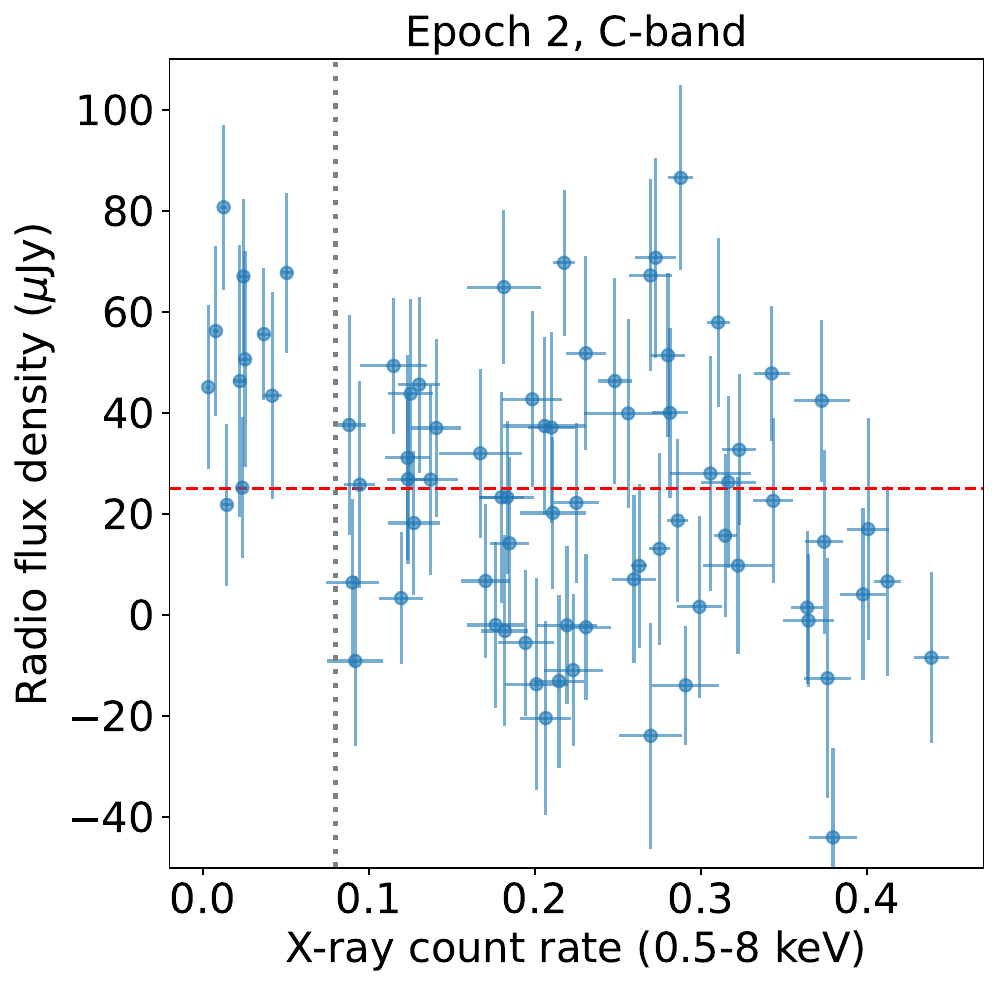}

\includegraphics[width=0.4\textwidth]{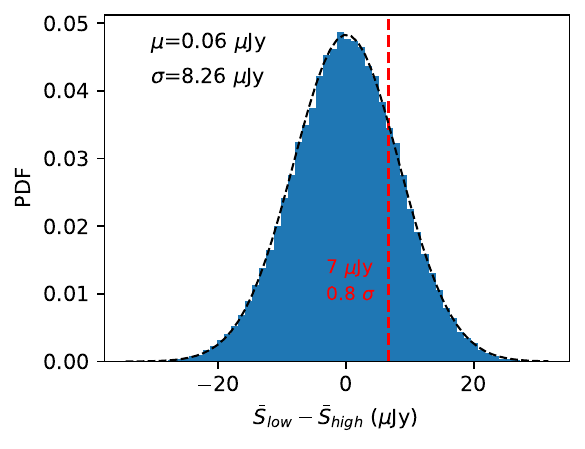}\hspace{5mm}
\includegraphics[width=0.4\textwidth]{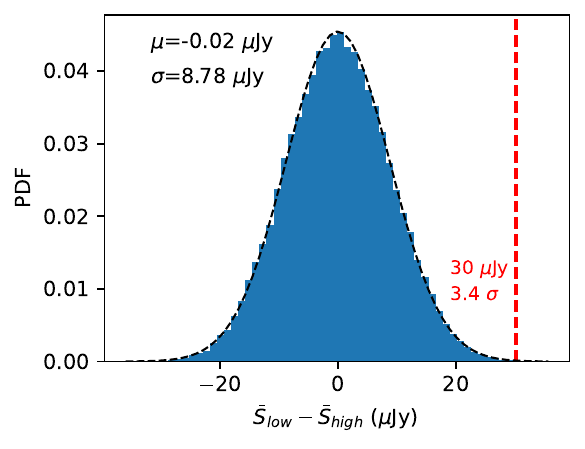}

\caption{{\it Top panels:} A scatter plot of radio flux density versus $0.5 - 8$\,keV averaged X-ray count rate for the strictly simultaneous VLA and {\it Chandra} observations of \fgl\ from Epoch~1 (left) and Epoch~2 (right). The horizontal red dashed lines corresponds to the average radio flux density of the source during the entire observation (in this case the mean of the radio lightcurve, which is consistent with the flux density obtained from the full-time image). The vertical grey dotted lines correspond to the averaged X-ray count rate that separates radio measurements associated with the low X-ray modes from the ones associated with the high X-ray modes (see Figure~\ref{fig:avx_cr_hist}). Note that for Epoch~1, points left of that line have, on average, the same radio flux density as the rest of the points. During Epoch~2, the radio flux density seems to be higher during the low X-ray modes compared to the rest of observation. {\it Bottom panels:} The histograms show the distribution of difference between the mean radio flux density in the low (points left of the grey dotted line in the top panels) and the high (points right of the grey dotted line) X-ray modes ($\bar{S}_{low} - \bar{S}_{high}$) taken from the sample of 10$^5$ randomly shuffled (in time) real radio lightcurves of \fgl\ for the Epoch~1 (left) and Epoch~2 (right). Measurements from the real (un-shuffled) radio lightcurves of \fgl\ are shown with the red vertical dashed lines. For the second epoch the actual \fgl\ measurement of $\bar{S}_{low} - \bar{S}_{high}$ is in a tail ($3.4 \sigma$) of the distribution, which means that the probability of radio emission of \fgl\ being brighter in the low X-ray modes just by chance is low.} \label{fig:r_vs_x_prob}
\end{figure*}
%%%%%%%%%%%%%%%%%%%%%%%%%

%Fig
%%%%%%%%%%%%%%%%%%%%%%%%%
\begin{figure*}
\centering
\includegraphics[width=0.4\textwidth]{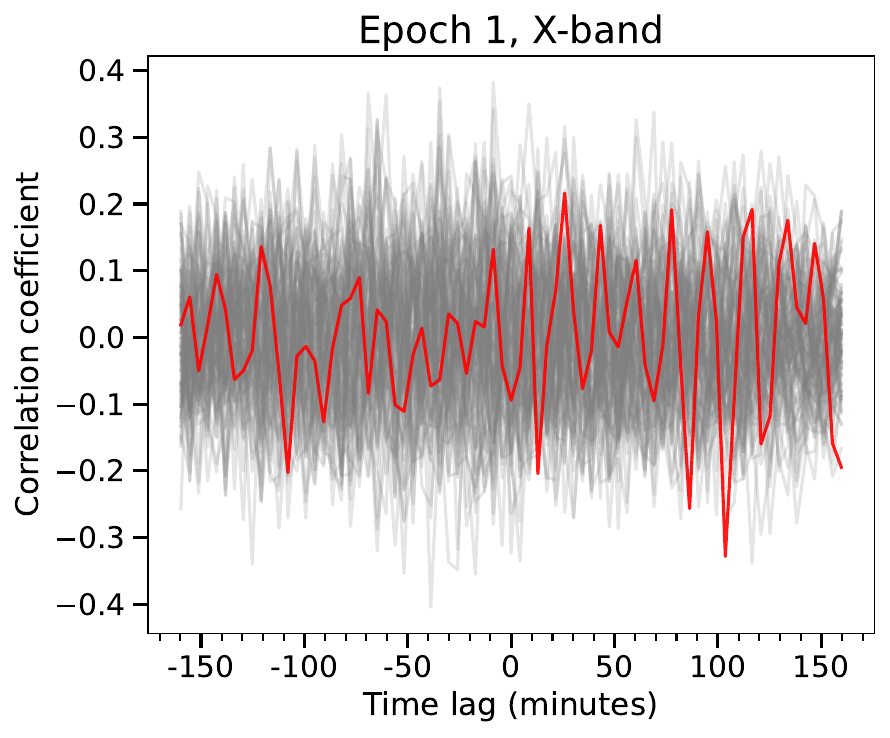}\hspace{5mm}
\includegraphics[width=0.4\textwidth]{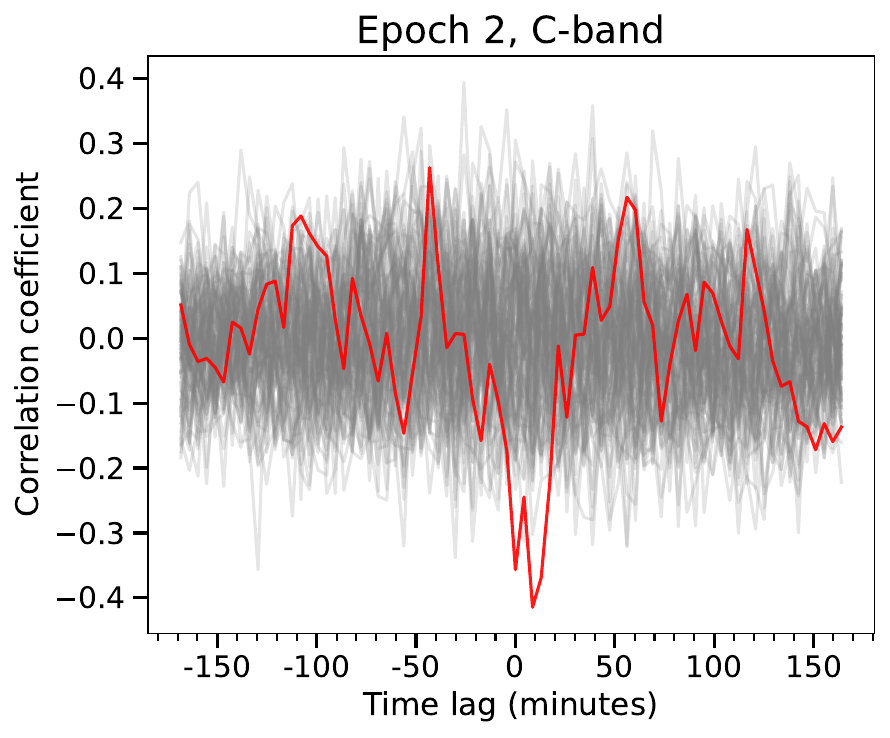}

\includegraphics[width=0.4\textwidth]{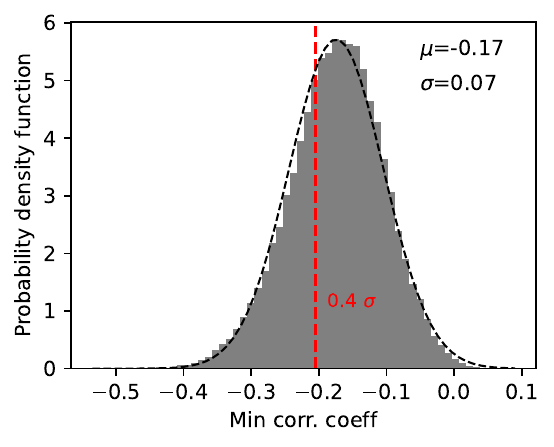}\hspace{5mm}
\includegraphics[width=0.4\textwidth]{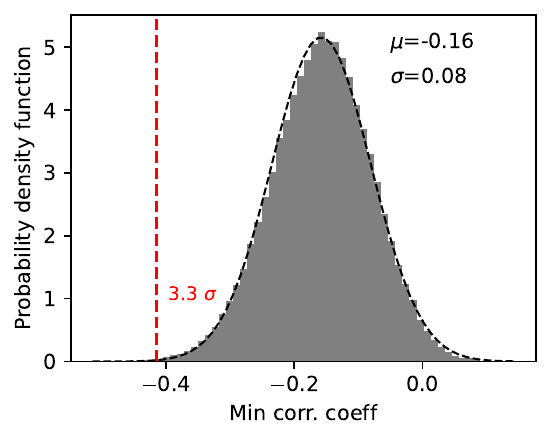}

\caption{{\it Top panels:} Correlation coefficient of the re-sampled $0.5 - 8$\,keV X-ray lightcurve of \fgl\ and its radio lightcurves from Epoch~1 (left) and Epoch~2 (right). The correlation of the real \fgl\ radio lightcurves are shown by the thick red line. The grey lines represent a sample of 100 correlation coefficient of reshuffled radio lightcurves of \fgl\ with its real X-ray lightcurve. {\it Bottom panels:} The histograms show the distribution of the minimal correlation coefficient near zero lag (interval between -50 and 50 minutes) of 10$^5$ randomly shuffled (in time) radio lightcurves of \fgl\ for the Epoch~1 (left) and Epoch~2 (right). Minimal correlation coefficient from the real (un-shuffled) radio lightcurves of \fgl\ are shown with the red vertical dashed lines. Anti-correlation of radio and X-ray lightcurves is detected in the second epoch with significance of 3.3 $\sigma$ (very similar to what we obtained by comparing radio fluxes in low and high X-ray modes; Figure~\ref{fig:r_vs_x_prob}).}\label{fig: cc_and_prob}
\end{figure*}
%%%%%%%%%%%%%%%%%%%%%%%%%

%Radio overall
\fgl\ is robustly detected in the radio during both epochs at a position consistent with the {\it Gaia} measurement ($15^{\rm h}44^{\rm m}39.^{\rm s}3888765\pm 0.^{\prime \prime}00018$; $-11^{\circ}28^{\prime}04.^{\prime \prime}87990\pm0.^{\prime \prime}00011$; \citealt{GaiaDR3}) and the previous best radio position of the source \citep{Jaodand2021}. In the first observation the source is found to be faint, with a flux density as low as $11.9\pm1.6\,\mu$Jy. The source is a factor of $2.5\times$ brighter during the second epoch with a mean flux density of $28\pm3\,\mu$Jy (see Table~\ref{tab:obs} and Figure~\ref{fig:radio_images}; note that these two observations are taken at different radio frequencies). 

\fgl\ is unresolved in both radio observations, and the first epoch (with the most extended A-configuration of the VLA and a synthesised beam of $0.41$\arcsec$\times$\ 0.25\arcsec) provides the best known radio position of the source (J2000):\\

\indent RA: $15^{\rm h}44^{\rm m}39.^{\rm s}389 \pm 0.^{\rm s}004\, (0.^{\prime \prime}06$.)\\
\indent Dec: $-11^{\circ}28^{\prime}04.^{\prime \prime}94 \pm 0.^{\prime \prime}03$.\\ 

We obtained this position using the {\tt imfit} task in {\tt CASA}. The fit errors represent the beam size divided by the S/N (not taking into account the absolute positional accuracy of the phase-referenced VLA observation). 

The low average radio flux density of \fgl\ imposes a substantial limitation on the shortest timescales that we were able to probe in our observations: longer averaging smooths the short-timescale variability of the source, and shorter averaging does not provide the sensitivity required for a detailed analysis. We created multiple radio lightcurves with time bins ranging from 1.75\,min up to an hour\footnote{12-min binned lightcurves are shown in Figure~\ref{fig:3sc_lcs_scaller} for reference.}, and found the optimal time binning (balancing sensitivity and averaging) to be $\sim3.5$\,min, which corresponds to individual radio scans of the source. The time-resolved lightcurves of \fgl\ for both epochs are shown in Figure~\ref{fig:radio_xray_lcs}. The average root mean square (RMS) noise in the individual images is 14\,$\mu$Jy and 20\,$\mu$Jy in the first and second epochs, respectively. Any intrinsic source variability would have to have a larger amplitude than that to be detectable. We found that, during both epochs, the radio flux density of \fgl\ was consistent within 1-$\sigma$ of the variability of other sources in the field (see Figure~\ref{fig:variability}) and even the background noise (see Figure~\ref{fig:background_stat}). Therefore, investigating the short-timescale behaviour of \fgl's radio emission and comparison with that of \jttt\ relies on a cross-correlation analysis of the source's radio and X-ray lightcurves.

%Fig
%%%%%%%%%%%%%%%%%%%%%%%%%
\begin{figure*}
\centering
\includegraphics[width=0.45\textwidth]{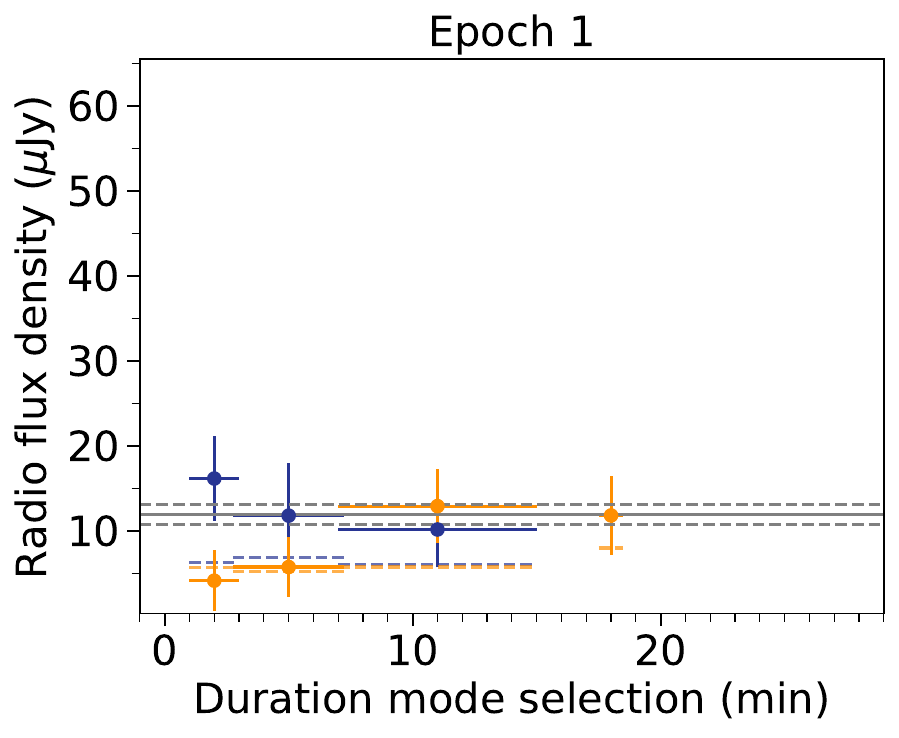}\hspace{10mm}
\includegraphics[width=0.45\textwidth]{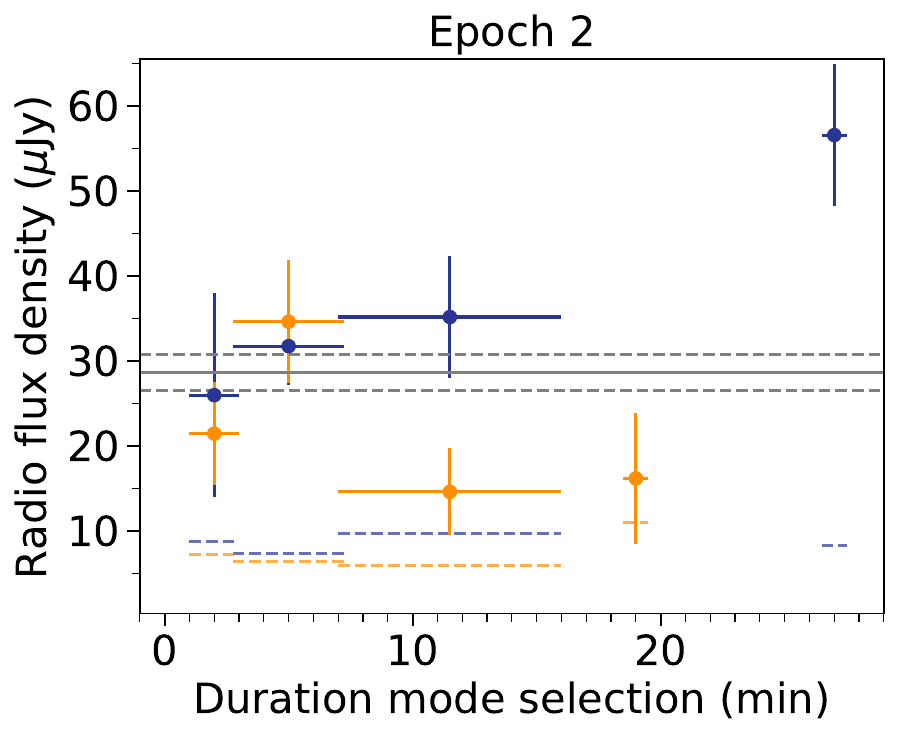}

\caption{Radio flux densities measured by averaging all low and high X-ray modes with corresponding interval of duration ({\it left}: Epoch~1; {\it right}: Epoch~2). The orange and dark blue points represent the measurements of the radio flux density in the low and high X-ray modes, respectively. Orange and blue horizontal dashed lines indicate the RMS of each image. The solid grey lines in each epoch correspond to the average flux density of \fgl\ over the entire epoch, with dashed lines indicating the 1$\sigma$ error. Note that the for the second epoch, the most significant difference of radio flux densities between two modes is observed when averaging modes that are longer than 5 minutes. During the first epoch, no significant differences are observed between low and high modes of any duration.}\label{fig:radio_modes_duraion}
\end{figure*}
%%%%%%%%%%%%%%%%%%%%%%%%%

\subsection{Anti-Correlation}\label{sec:anti-correlation}

Despite the observed intrinsic variability of \fgl\ being close to the RMS noise levels, if \fgl\ exhibits the same anti-correlated radio and X-ray behaviour as \jttt\ \citep{Bogdanov2018}, we should still be able to see evidence of it in cross-correlation analysis between the two lightcurves (see comparative analysis of \jttt\ and \fgl\ in Section~\ref{sec:j1023_compare}).

The X-ray emission of \fgl\ is observed to modulate on timescales as short as 30\,s (Figure~\ref{fig:xray_lcs}), while we cannot achieve shorter than $\sim3.5$\,min time resolution in radio and still provide adequate sensitivity for a detailed analysis. Thus, we approach the correlation analysis in two ways: 1) ``Evenly spaced lightcurve approach'' -- adjusting the X-ray binning to match the radio bins and 2) ``X-ray modes binning approach''  -- imaging time-windows of radio data to exactly span the time intervals of X-ray modes.

\subsubsection{Evenly spaced lightcurves}\label{sec:ac_even}

Following the first approach, we averaged the X-ray count rate inside each $\sim3.5$-min radio time bin for both epochs. The resulting distribution of average X-ray count rates per radio scan is shown in Figure~\ref{fig:avx_cr_hist}. Despite the significant downsampling (by a factor of 7), a visible separation of low-mode and high-mode X-ray bins is still observed at an average count rate of 0.08\,cts~s$^{-1}$. Creating X-ray and radio lightcurves with equal time-binning allowed us to compare the radio and X-ray fluxes directly. 

The top panels of Figure~\ref{fig:r_vs_x_prob} present a scatter plot of the resulting coupled radio and X-ray flux measurements of \fgl. We found no signs of correlation between the X-ray and radio emissions of \fgl\, during the first epoch.
In contrast, the second epoch reveals a hint of anti-correlation. This is noticeable in the bottom panel of Figure~\ref{fig:radio_xray_lcs}, where radio flux appears to interchange with X-ray flux. The scatter plot (Figure~\ref{fig:r_vs_x_prob}, top right) further emphasises this, as points corresponding to low X-ray count rate show, on average, higher radio flux density compared to the rest of the points and the mean flux density of \fgl\ (obtained by imaging the entire time of the second observation). 

To estimate the statistical significance of the observed difference between radio flux density in low and high X-ray modes in \fgl, we considered the difference between the mean radio flux density $\bar{S}_{low}$ of the 9 (Epoch 1) and 11 (Epoch 2) points that coincided with X-ray low modes (points left of the grey dotted line in the top panels of Figure~\ref{fig:r_vs_x_prob}) and the mean radio flux density $\bar{S}_{high}$ of the rest of the radio light curve (points right of the dotted grey line in the top panels of Figure~\ref{fig:r_vs_x_prob}). We found that $\bar{S}_{low} - \bar{S}_{high}$ is $\sim 7\, \mu$Jy in the first epoch and $\sim 29\, \mu$Jy in the second epoch, while the standard deviation of the entire lightcurve of \fgl\ is $15\, \mu$Jy and $26\, \mu$Jy in the first and second epoch respectively. To estimate how likely it is for all the radio flux density measurements that happen to have low corresponding X-ray count rate to be statistically brighter, we generated a set of 10$^5$ radio lightcurves, each a copy of the original \fgl\ lightcurve but with radio flux measurements shuffled randomly in time (using {\tt np.random.shuffle}). We computed $\bar{S}_{low} - \bar{S}_{high}$ for each of these lightcurves and plotted the resulting distribution in the bottom panels of Figure~\ref{fig:r_vs_x_prob}. The mean and standard deviation of the resulting distributions of $\bar{S}_{low} - \bar{S}_{high}$ are $\mu=-0.05,\, \sigma=8.26\, \mu$Jy and $\mu=0.0,\, \sigma=8.8\, \mu$Jy for the first and second epoch, respectively. This puts values measured from the real lightcurves of \fgl\ above the mean by $0.8\, \sigma$ in the first epoch and $3.4\, \sigma$ in the second epoch. Thus, at least in the second epoch, the radio emission of \fgl\ seems to be higher in low X-ray modes, and it is unlikely to be by chance.

We also performed a direct correlation of radio lightcurves of \fgl\ and its re-sampled X-ray lightcurves (see Figure~\ref{fig: cc_and_prob}). The correlation coefficient was calculated using the {\tt stingray} python package, and scaled as follows: 

$$C_c = \frac{C_{XR}}{\sqrt{A_{X} \times A_{R}}}$$

Here, $C_{XR}$ represents the cross-correlation between X-ray and radio lightcurves, and $A_{X}$ and $A_{R}$ denote the maximum absolute values of the auto-correlations of X-ray and radio lightcurves, respectively.

Similarly to the previous analysis, the first epoch revealed no relation between radio and X-ray lightcurves, while a visible anti-correlation is revealed in the second epoch. In the second epoch, the absolute value of the negative correlation coefficient near zero lag exceeds that of the majority of randomly shuffled radio lightcurves (giving a $\sim$3.3 $\sigma$ significance). This further confirms our detection of a short-timescale anti-correlation of radio and X-ray emission in \fgl.

The detected anti-correlated radio and X-ray variability in \fgl\ is unlikely to be an artefact of the residual variability noise of the radio images because both the minimum correlation coefficient near zero lag and the $\bar{S}_{low} - \bar{S}_{high}$ values measured from the second epoch of \fgl\ far exceed measurements obtained from lightcurves of $\sim$20 radio sources from the nearby field of \fgl\ image (made from the same images of the same radio scan time bins as \fgl's lightcurves; see Appendix~\ref{sec:appedix_var}, Figure~\ref{fig:variability}).

\subsubsection{X-ray modes binning}

Following the second approach, we first analysed radio images created by integrating the time intervals corresponding to individual low and high X-ray modes (longer than $\sim$3.5 minutes). We did not detect any significant differences between the two groups of radio flux density measurements (associated with low and high X-ray modes) likely due to low signal to noise ratio (SNR) of individual measurements.

To increase the S/N, we created radio images by stacking all the times when \fgl\ was in low or high X-ray mode. Radio flux densities of \fgl\ obtained by averaging together all (longer than 3.5\,min) high X-ray modes and (in a separate image) low X-ray modes were measured to be $S_{high}=10\pm4\, \mu$Jy, $S_{low}=10 \pm 4\, \mu$Jy for the first epoch and $S_{high}=22\pm 4\, \mu$Jy, $S_{low}=41 \pm 6\, \mu$Jy for the second epoch. From this, we find no difference between the two modes in the first epoch, but a significant difference is observed in the second epoch, consistent with the previous results. 

We further explored the radio data by imaging together the modes that have a similar duration. The specific duration intervals for those combinations are chosen based on the distribution of mode duration in the observed X-ray lightcurves of \fgl\ (see Figure~\ref{fig:xraymodes}, where vertical lines indicate the chosen intervals). We find variability of the radio flux density of \fgl\ by a factor of a few in different windows and modes (see Figure~\ref{fig:radio_modes_duraion}), although in the case of the first epoch, this variability is mostly contained within the error-bars.
In the second epoch, the longest modes exhibit the most significant difference of their radio flux density.
This behaviour of \fgl\ during the second epoch is consistent with what is seen in \jttt, where the short low X-ray modes were observed to have lower associated radio flux densities than the longer ones (see Figure~4 of \citealt{Bogdanov2018}).

\begin{figure*}
\centering
\includegraphics[width=\textwidth]{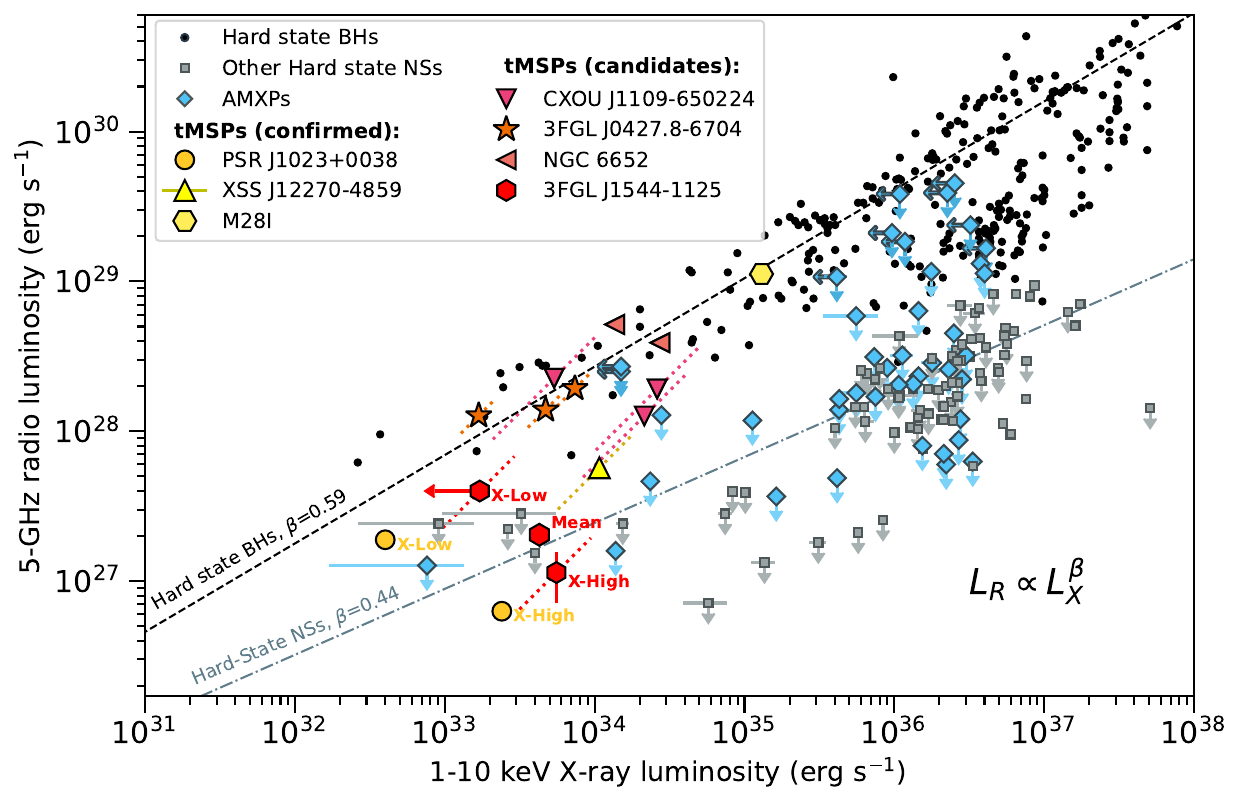}

\caption{(Quasi-)simultaneous radio and X-ray luminosities for BH- and NS-LMXBs. Black circles represent BH-LMXBs; grey squares represent NS-LMXBs; light blue diamonds represent AMXPs; various yellow symbols show 3 confirmed tMSPs; various symbols of red hue show tMSP candidates; Our simultaneous Radio---X-ray measurements of \fgl\ are shown using red hexagons (where the assumed distance is 3.45\,kpc; and only measurements from Epoch 2 are plotted: the average luminosity as well as the luminosities during the longest low and high X-ray modes). For each tMSP and tMSP candidates we also indicated the range of possible luminosities caused by the large uncertainties on the distances to these sources (dotted diagonal lines of the corresponding color). Data points are taken from \citet{Bahramian2018} for hard state BHs; \citet{MIGFEN2006,Tetarenko2016,Gusinskaia2017,Gusinskaia2019} for hard state neutron stars; \citet{CotiZelati2019} for M28I; \citet{Hill2011} for XSS~J12270$-$4859; \citet{Bogdanov2018} for PSR~J1023+0038; \citet{Tudor2017}, \citet{Tetarenko2018}, \citet{Tudor2016ATel}, \citet{MJ2010ATel}, \citet{Tetarenko2017ATel}, \citet{Migliari2011}, \citet{J17379_ATel1148},\citet{Russell2018} and \citet{Gusinskaia2019} for AMXPs; \citet{CotiZelati2021} for CXOU~J1109$-$650224; \citet{Li2020} for 3FGL~J0427.8$-$6704; \citet{Paduano2021} for NGC~6652B. Correlation tracks for hard state BHs (dashed black line) and NSs (dash-dotted grey line) are defined in \citet{Gallo2018}.}\label{fig:lrlx}
\end{figure*}

We used the second epoch radio flux density measurements of the longest low X-ray mode and the longest high X-ray mode to estimate the mode-separated luminosity of the source. These are the fluxes and luminosities quoted in Table~\ref{tab:obs} and shown in Figure~\ref{fig:lrlx}.

\subsection{Radio spectrum}

The difference in the mean radio flux density of \fgl\ between the first and the second epoch raises the question of whether the source's spectral index could be responsible for the observed brightening. Simply taking the mean radio flux density value at each epoch, we measure a spectral index of $-2.1\pm0.3$, exceptionally steep for an accretion-powered source (although not unusual for a radio pulsar), and much steeper than seen in \jttt\ ($-0.5<\alpha<0.4$; \citealt{DEL2015,Bogdanov2018}). To cross-check this, we attempted to measure the spectral index within Epoch~2, as the C-band data spanned $4-8$\,GHz. We obtained a consistent spectral index of $\sim=-1.9\pm0.8$. We also attempted X-ray mode separated spectral index measurements, both between epochs and within the second epoch. The results of all our spectral index measurements are listed in  Table~\ref{tab:radio_spectrum} and discussed in Section \ref{sec:disc_radio_spec}. 

%Tab2
%%%%%%%%%%%%%%%%
\begin{table}
\caption[]{Radio spectral index measurements of \fgl}

\begin{minipage}{\textwidth}
\renewcommand{\arraystretch}{1.2}
{
\begin{tabular}{@{\extracolsep{-3pt}}l|cccc@{}}
\hline\hline
Data & $\nu$ (GHz) & $\alpha$ \\
\hline
Ep 1, Ep 2 (entire observation) & 6.1, 9.5  & $-2.1\pm0.3$\\
Ep 2 (entire observation) & 4.8, 5.7, 6.6, 7.5 & $-1.9\pm0.8$\\
\hline
Ep 1, Ep 2 (long low modes) & 6.1, 9.5 & $<-2.8$\\
Ep 2 (longest low mode) & 4.8, 5.7, 6.6, 7.5 & $-1.8\pm0.8$\\
Ep 2 (low X-ray flux)$^{a}$ & 4.8, 5.7, 6.6, 7.5 & $0.5\pm0.8$\\
\hline
Ep 1, Ep 2 (long high modes) & 6.1, 9.5 & $-0.7\pm0.8$\\
\hline\hline
\end{tabular}
}
\begin{flushleft}{
The $\nu$ column lists the central frequency of each measurement\\
used for spectral index estimation.

$^{a}$ Points left of the grey dotted line in Figure~\ref{fig:r_vs_x_prob} (top right).
  }\end{flushleft}
\end{minipage}
\label{tab:radio_spectrum}
\end{table}
%%%%%%%%%%%%%%%

%Fig
%%%%%%%%%%%%%%%%%%%%%%%%%
%%%%%%%%%%%%%%%%%%%%%%%%%

\subsection{Position on \lr-\lx}

To assess the radio and X-ray behaviour of \fgl\ in comparison to other accreting systems, we need to convert its observed fluxes into luminosities. Currently, there are two available distance measurements for the source: 1) the `photometry' distance of 3.8$\pm$0.7\,kpc (\citealt{Britt2017}; obtained from the estimation of the absolute magnitude of the donor star from its spectral properties) and 2) the parallax-based distance of 3.1$\pm$0.5\,kpc \citet{BailerJones2021}; obtained using the geometrical parallax and optical magnitude of \fgl\ reported in {\it Gaia} DR3 \citep{GaiaDR3} and refining the distance via Galactic 3D modelling. These two measurements are consistent with each other, and we will use the average value of 3.45\,kpc for \fgl\ luminosity estimations but take into account the full range of the errors for each measurement.

For a distance of 3.45\,kpc, the unabsorbed X-ray $0.5-8$\,keV flux during high and low modes translates to $1-10$\,keV luminosities of \lx$_{low} < 1.7\times 10^{33}$\,\ergs, \lx$\,_{high} = (5.5\pm 0.1) \times 10^{33}$\,\ergs\ for both epochs (see Table~\ref{tab:obs}). These luminosities are comparable to high-mode luminosities of \jttt\ and other tMSP candidates in the LMXB-like state. Given differences in radio flux densities between the two epochs, for comparison with other systems on the \lr/\lx\ diagram we use only the second epoch measurements since they are done at C-band ($4-8$\,GHz), which matches closer to the diagram's chosen reference frequency, and they show moding. The radio flux densities of \fgl\ translate to 3.45-kpc luminosities (\lr$=4 \pi D^2 \nu S_{\nu}$) of \lr$\,_{low} = (4.0\pm 0.6) \times 10^{27}$\,\ergs, \lr$\,_{high} = (1.1\pm 0.4) \times 10^{27}$\,\ergs. The results are plotted in Figure~\ref{fig:lrlx}, which shows that \fgl\ occupies a similar region in the \lr/\lx\ diagram as  tMSPs and other tMSP candidates, closely matching the measurements of \jttt.

%Fig
%%%%%%%%%%%%%%%%%%%%%%%%%
\begin{figure*}
\centering
\includegraphics[width=0.9\textwidth]{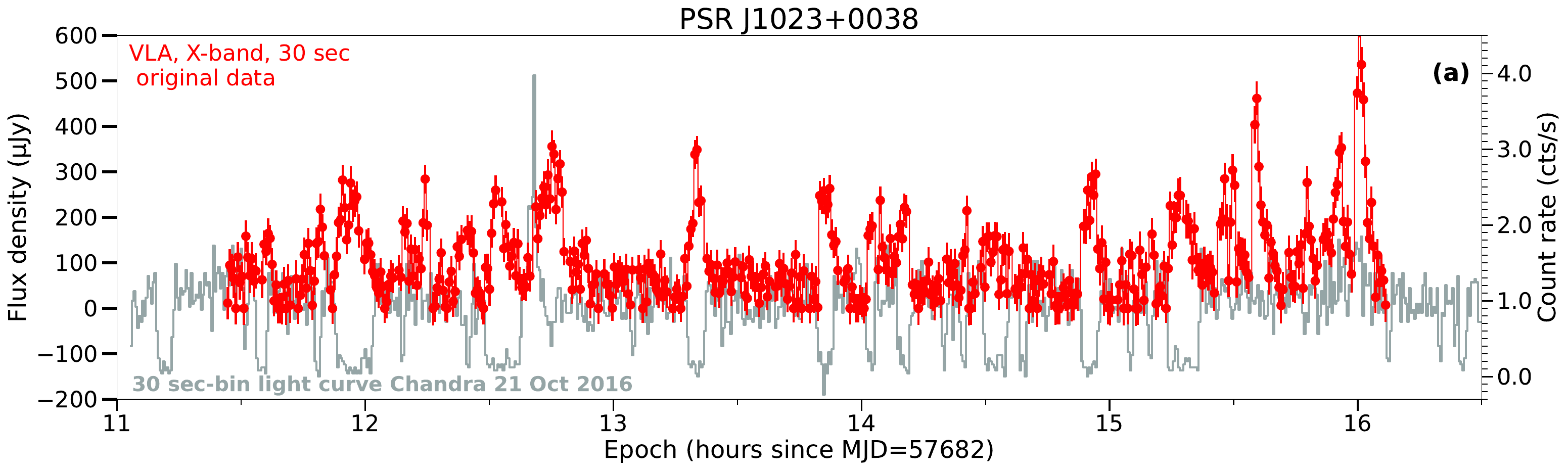}
\includegraphics[width=0.9\textwidth]{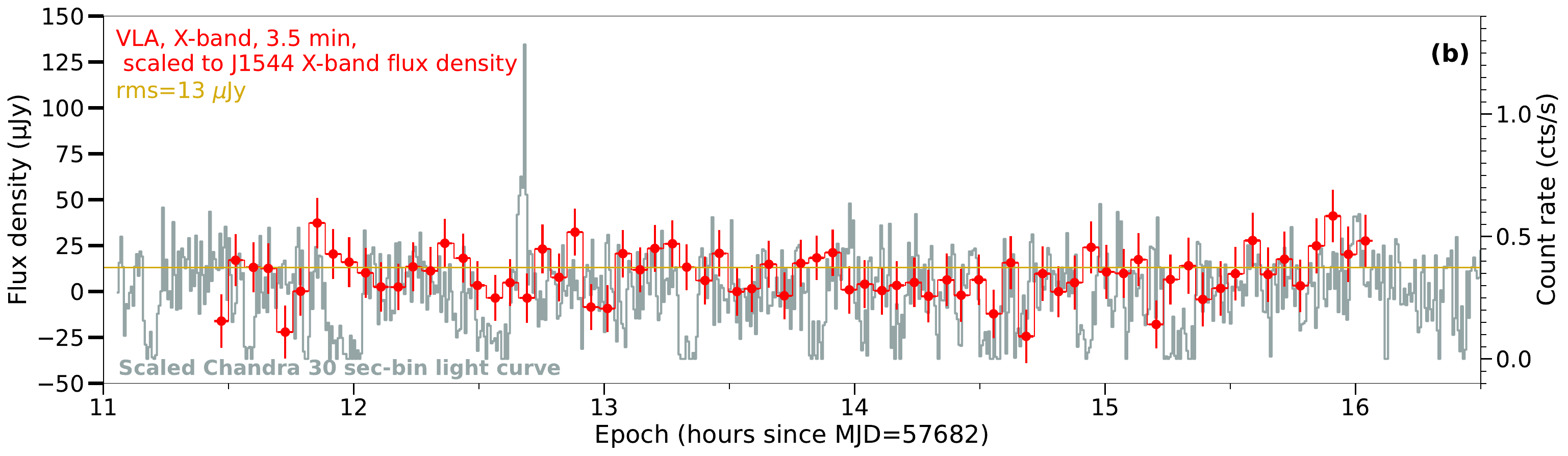}
\includegraphics[width=0.9\textwidth]{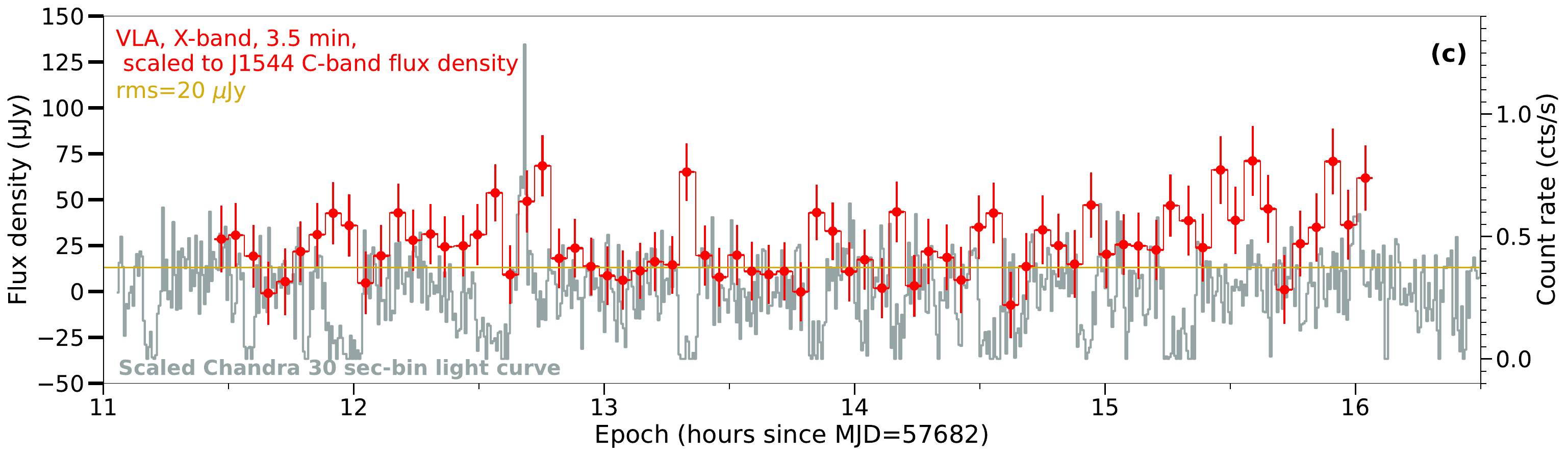}

\includegraphics[width=0.43\textwidth]{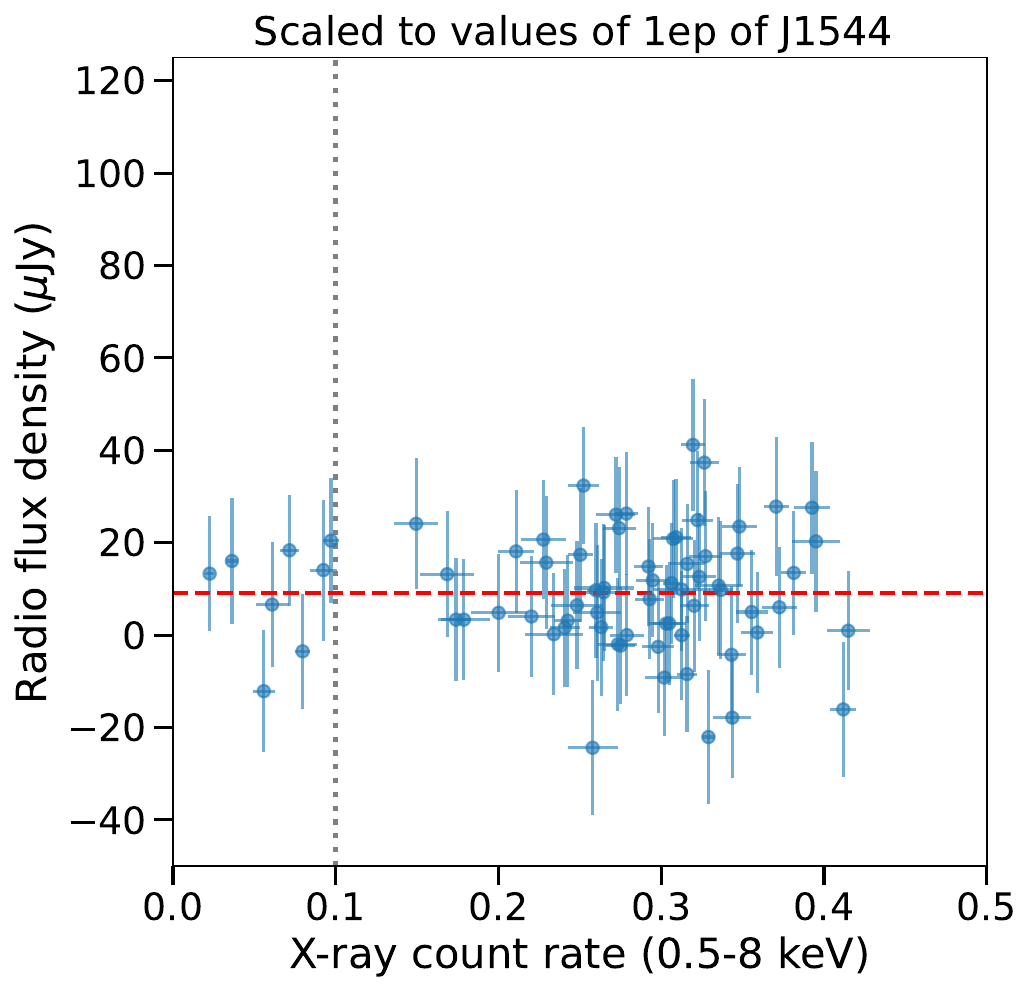}\hspace{5mm}
\includegraphics[width=0.43\textwidth]{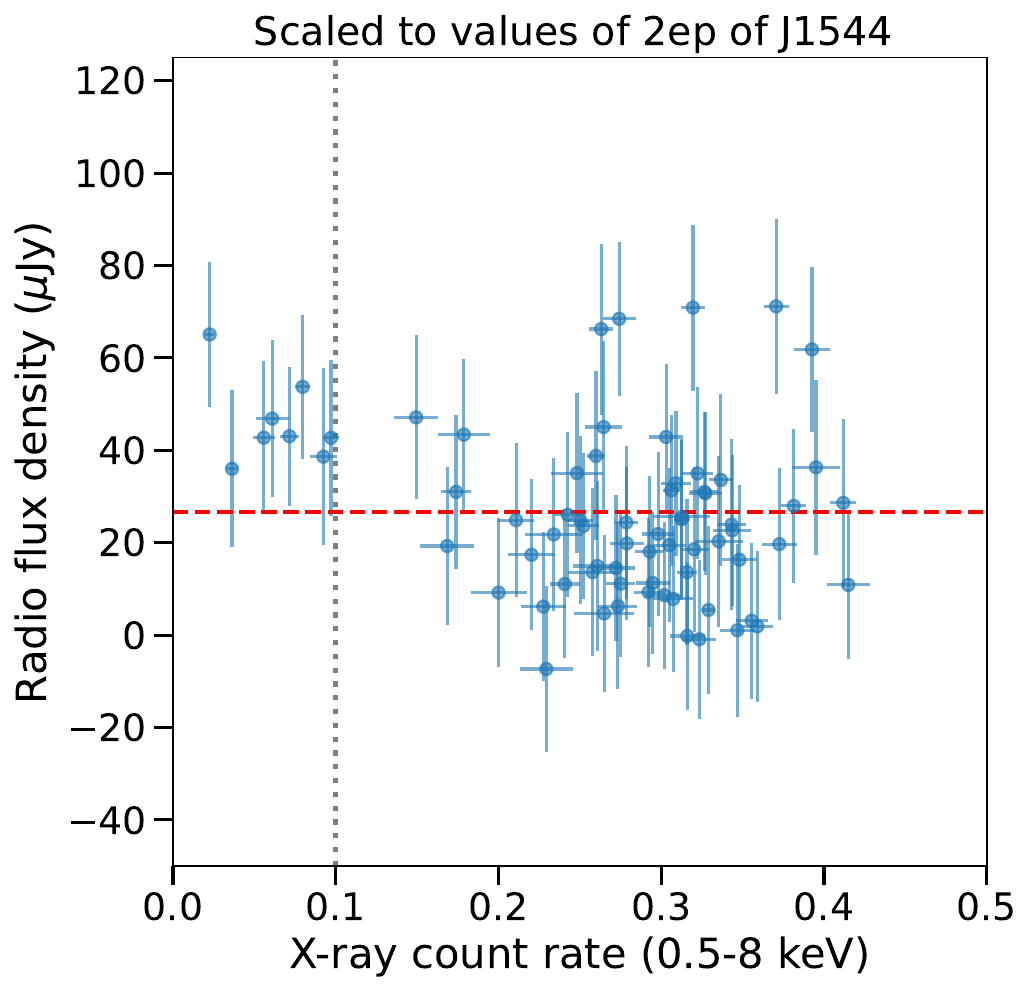}

\caption{{\it Top panels:} VLA and {\it Chandra} lightcurves of \jttt\ taken from \citealt{Bogdanov2018}; {\bf (a)} the original lightcurves with a native sampling of 30 sec and original flux levels. {\bf (b)}: the same lightcurves but scaled to X-ray count rate ($\sim$0.3) and radio flux density of \fgl\, during the first epoch of this campaign ($S_{\nu}\approx 12 \mu$Jy) and with added noise. The radio lightcurve is also down-sampled by a factor of 7. {\bf (c)}: same as {\bf b}, but scaled to the fluxes of \fgl\, during the second epoch of this campaign ($S_{\nu}\approx 28 \mu$Jy).
{\it Bottom panels:} Scatter plots of radio flux density versus averaged X-ray count rate obtained from lightcurves of panel {\bf b} - left, and {\bf c} - right. Note similarity with top panels of Figure~\ref{fig:r_vs_x_prob}.}\label{fig:J1023_compare}
\end{figure*}
%%%%%%%%%%%%%%%%%%%%%%%%%

\subsection{Comparison to PSR J1023+0038}\label{sec:j1023_compare}

\jttt\ exhibits a very clear anti-correlation between X-ray and radio fluxes (when excluding flaring-mode data; \citealt{Bogdanov2018, Paduano2021}), while the observed hint of a similar anti-correlation in \fgl\ is less convincing due to the larger distance to the latter. Our findings raise the question of whether \fgl\ behaves differently from \jttt\ or if it's simply challenging to capture potential radio emission variability at the flux levels of \fgl. To answer this question, we simulated placing \jttt\ at larger distance ($\sim2.5$\,kpc) to match its X-ray and radio fluxes to those of \fgl.

We took {\it Chandra} and VLA 30-s binned lightcurves of \jttt\ (shown in Figure~\ref{fig:J1023_compare}, panel {\bf a}; produced from data from the 2016 campaign reported by \citealt{Bogdanov2018}) and scaled its values by a factor of $\sim4$ to match the mean X-ray count rate and the mean radio flux densities of \fgl's first and second epoch measurements of the campaign presented here. To further improve the comparison, the radio lightcurve was down-sampled by a factor of 7 to have the same time resolution as that of \fgl. We also added noise to the resulting X-ray and radio lightcurves that corresponds to an average background noise for 30-s and 3.5-min integrations in {\it Chandra} and VLA data, respectively. The resulting scaled X-ray and radio lightcurves of \jttt\ are shown in Figure~\ref{fig:J1023_compare}, panels ({\bf b}-{\bf c}). These lightcurves closely resemble those of \fgl\ and the high noise level masks the anti-correlated variability almost entirely, especially for the lower flux X-band observation. The scatter plots (shown at the bottom panels of Figure~\ref{fig:J1023_compare}) don't reveal any anti-correlation in the lightcurves scaled to the first epoch fluxes of \fgl, and suggest a hint of anti-correlation in the lightcurves scaled to the second epoch fluxes. In fact, the striking similarity between Figure~\ref{fig:r_vs_x_prob} and Figure~\ref{fig:J1023_compare} indicates that \fgl\ appears to behave very similarly to \jttt. An increased sensitivity of the radio measurements, by a factor of at least a few, is necessary to further study \fgl\ in detail.

\section{Discussion}
\label{sec:discussion}

\jttt\ is is the first-discovered and the most-studied tMSP in the low-luminosity accretion disk state. Due to the scarcity of confirmed tMSP systems, there is a tendency to extrapolate phenomena observed solely in \jttt\ to the tMSPs as a class. Importantly, such extrapolation should always be supported by evidence of similar behaviours from other sources that belong to a class of tMSPs or tMSP candidates. In this study, we investigate the short-timescale relationship between X-ray and radio emission in a very strong tMSP candidate system, \fgl. 

\fgl\ is a strong tMSP candidate due to a set of its properties that closely resemble those of the known tMSP population (\citealt{Bogdanov2015}, and see Section~\ref{sec:into-j1544}). Currently \fgl\ resides in the low-luminosity LMXB-like state (\lx$\sim10^{33}$\,erg~s$^{-1}$), where it displays bi-stable X-ray moding. It is an active radio source with luminosity and variability matching those of known tMSPs \citep{Jaodand2021}. 

\subsection{Radio and X-ray behaviour of \fgl}

Our dual-epoch simultaneous VLA and {\it Chandra} campaign revealed that during both observations \fgl\ showed behaviour in both X-ray and radio bands that was consistent with previous observations of this system. The X-ray lightcurve was observed to switch between low and high X-ray flux modes on timescales of minutes (with no obvious X-ray flare events), spending $\sim$20\% of the time in the low X-ray mode. The measured X-ray fluxes of $\sim2-4\times10^{-12}$\,erg ~cm$^{-2}$~s$^{-1}$ and X-ray power-law spectrum with $\Gamma \approx 1.7$ were found to be almost identical to what was observed in \fgl\ in 2014 and 2015 \citep{BogHal2015, Jaodand2021}. 

Our measurements of the radio flux density of \fgl\ in both X- and C-bands ($\sim 11-56\, \mu$Jy) are consistent with results from previous radio campaigns on this source \citep{Jaodand2021}.\footnote{The $\gamma$-ray counterpart of 3FGL~J1544.6$-$1125 has also been the target of a short ATCA observation as a part of the campaign of \citet{Petrov2013}. However, the 5.5-GHz flux density of $10\pm0.6$\,mJy listed for \fgl\ arises from a nearby confusing source, which has coordinates that are consistent with source `R4' in our field (see Figure~\ref{fig:radio_images}).} We find significant differences in the radio flux density of \fgl\ between the two observations of our campaign. These differences could be attributed to (i) spectral properties of \fgl's emission (as the two observations recorded data at different frequency bands), (ii) intrinsic long-term variability of the source (as the observations were performed 8 months apart), or (iii) similar to \jttt\, it could be related to the short-term variability of radio emission connected to X-ray moding in the system, which is unique in each observation. Below we explore each possibility. 

\subsubsection{Long-term radio variability}\label{sec:long_term_var}

Quasi-simultaneous VLA (10\,GHz) and {\it Swift}-XRT observations of \fgl\ in 2015 revealed that the system was varying by a factor of $2-3$ between epochs separated by a few weeks with flux densities of $<$13.8, $<$16.2, 23.6 and 47.7\,$\mu$Jy. Short-timescale ($\sim3$-min bin) lightcurves of \fgl\ in 2015 (see Figure~3 in \citealt{Jaodand2021}) seem to indicate that the source is variable on longer than 40-min timescales. In its brightest epoch, the flux density stayed above $50\,\mu$Jy for more than half of the 40-min observation, while in its faintest epoch, the source was undetected in each of ten 5-min light-curve bins (neither in full-time integration). Similar to the faintest epoch of 2015, during the majority of the 2018 X-band observation of this campaign (the same band as the 2015 campaign), \fgl\ stayed below the RMS noise of $\sim13\,\mu$Jy with only 2 points above $S_{\nu}\sim50\,\mu$Jy (see Figure~\ref{fig:radio_xray_lcs}). In our 2019 C-band observation, \fgl\ was brighter overall in the radio band, with a continuous period of $\sim$40\,min for which its flux density stayed above $S_{\nu}\sim50\,\mu$Jy. Moreover, the source seems to stay brighter during the entire C-band observation, compared to the entire time in the X-band observation (even excluding radio points associated with the longest low X-ray mode in the former). Thus, it is entirely plausible that underlying long-term variability is responsible for the differences in average radio flux density of individual radio epochs of \fgl, which is not directly related to the X-ray moding in the system. It cannot be related to the long-term X-ray variability of \fgl\ as the two X-ray observations of our campaign revealed almost identical fluxes and spectra (see Section~\ref{sec:res-xray}).

To search for this potential long-term radio variability in \fgl, more observations with higher cadence will be required. Those would have to be performed in the same radio frequency band and accompanied by strictly-simultaneous X-ray observations to rule-out spectral or X-ray mode-driven differences. Simultaneous observations in other bands may also be helpful to test if long-term radio variability could be related to, for example, optical flaring in \fgl\ \citep{Bogdanov2015}.

\subsubsection{Anti-correlation of X-ray and radio fluxes}

We found a hint of anti-correlation between radio and X-ray fluxes in our second (C-band) observational epoch of \fgl. The radio flux density appears to be on average brighter when the X-ray flux of the source is low (below a {\it Chandra} count-rate of $\sim0.1$\,cts~s$^{-1}$; see Figure~\ref{fig:r_vs_x_prob}). This anti-correlation is also visible in the direct cross-correlation analysis (see Figure~\ref{fig: cc_and_prob}). Comparison with the distribution of randomly shuffled radio light curves of this observation gave a significance of this finding of 3.4$\sigma$ (see Section~\ref{sec:ac_even}). This behaviour is consistent with the anti-correlated radio and X-ray emission behaviour observed in the low X-ray modes of \jttt. In fact, if we scale and down-sample \jttt\ fluxes to the level of those in \fgl, the very clear anti-correlated behaviour of \jttt\ light curves appears at the same level seen in both epochs of this campaign (see Figure~\ref{fig:J1023_compare} and Section~\ref{sec:j1023_compare}). 

During our observations, \fgl\ exhibited four low X-ray modes that are longer than 10 minutes (the longest one being 28 min, in the second epoch), while all low X-ray modes in the 2016 observation of \jttt\ \citep{Bogdanov2018} were shorter than 10 min. This perhaps played a role in our ability to detect the anti-correlated radio and X-ray emission in the faint \fgl, because, similarly to \jttt, \fgl\ seems to be brighter in the long X-ray modes than in short ones (see Figure~\ref{fig:radio_modes_duraion}).  We also see that the radio flux density appears to be lower in the long X-ray high modes than in shorter ones. This could indicate that the mechanism responsible for the emission in both modes requires some time to fully switch to one state to another. However, given the very small number of observed modes and the low SNR it is hard to make any solid conclusion regarding this.

We note that with the sensitivity and time resolution of \fgl's light curves, it is impossible to take into account effects of any potential radio flares that are not associated with the X-ray low modes, which are a common occurrence in \jttt. In the 2016 observation of \jttt, an obvious radio flare was observed at the same time as the only X-ray flare of the observation (occurring at around 12.5\,hr; see the top panel of Figure~\ref{fig:J1023_compare}). There are also multiple radio flares observed at the end of \jttt's observation that occur during the high X-ray mode. These modes are completely indistinguishable from the general high X-ray mode-associated radio emission of \jttt\ in the scaled and down-sampled versions of the light curves (see panels {\bf b}-{\bf c} of Figure~\ref{fig:J1023_compare}). The presence of these flares precludes the direct detection of an anti-correlation between the radio and X-ray lightcurves via cross-correlation analysis. \citet{Paduano2021} showed that the anti-correlation was only detected in \jttt\ when flaring-mode data were excluded. In contrast, we tentatively detect the anti-correlation of \fgl's radio and X-ray lightcurves without exclusion of any data. If \fgl\ behaves in a similar way to \jttt, a marginal anti-correlation seen in the correlation coefficient curve of \fgl\ might indicate the absence of radio flares in the system. Indeed, \fgl\ does not exhibit X-ray flaring. However, a suggestion of `orphan' radio flares can be seen in both \fgl\ observations of this campaign as measurements of bright radio flux densities at high X-ray count rates (see Figure~\ref{fig:r_vs_x_prob}; and perhaps even more obviously in the radio lightcurves with longer radio time bins shown in Figure~\ref{fig:3sc_lcs_scaller}). The presence of these potential flares complicates the comparison of radio flux densities associated with the high and low X-ray modes of \fgl.

The comparative analysis of \jttt\ also gives an answer to why we see hints of anti-correlation only in the second epoch: for the same average flux density as that of the first epoch of \fgl\ ($12$\,$\mu$Jy), the moding behaviour of \jttt\ would have been impossible to detect due to low S/N. In the reverse case, if \fgl\ was as bright at X-band as it was at C-band, we would have been able to detect similar hints of its anti-correlated behaviour. \fgl\ was previously observed to be as bright as $40$\,$\mu$Jy at X-band \citep{Jaodand2021}, thus, if there is long-term variability of the brightness, future attempts at capturing anti-correlated radio/X-ray behaviour in this source could be more successful. 

\subsubsection{Spectral properties}\label{sec:disc_radio_spec}

As discussed above, the significant difference in radio brightness of \fgl\ between the two separate observations of this campaign could be caused by intrinsic spectral properties of the source. 
Indeed, the measured very steep radio spectral index $\alpha\sim-2$ is consistent between the following two methods (see Table~\ref{tab:radio_spectrum}): 1) comparing flux densities between two epochs ($\nu\sim$6.1 and 9.5\,GHz) and 2) the splitting the second epoch in 4 sub-bands ($4<\nu<8$\,GHz). This is inconsistent with what is observed in \jttt\ \citep{DEL2015,Bogdanov2018}, which showed a flat spectral index on average.

We tentatively find that during high X-ray modes the radio spectral index of \fgl\ was consistent with being flat or steep ($\alpha=-0.7\pm0.8$), and in the long low X-ray modes it was consistent with being very steep ($\alpha<-2$). However, the average of all low-X-ray-flux time windows of Epoch 2 gives an inconsistent measurement of $\alpha=0.3\pm0.8$. This could indicate that shorter low-modes exhibit a significantly different (perhaps even inverted) spectrum or, which is more likely, this indicates that intraband spectral index measurements of very faint sources such as \fgl\ are unreliable. Thus, due to the poor sensitivity of our measurements and some inconsistency in measurements from different approaches (see Table~\ref{tab:radio_spectrum}), we are unable to draw definitive conclusions about the spectral variability of \fgl using our current data.

Nevertheless, we cannot ignore the difference in the overall radio brightness of \fgl\ between the two epochs. If these differences are indeed due to an intrinsic steep radio spectrum, it will be worth repeating the same campaign at lower radio bands, such as with the S-band ($2-4$\,GHz) receiver of VLA or MeerKAT. In fact, there is an indication that the radio emission of the other strong tMSP candidate CXOU~J110926.4$-$65022 can also have a steep radio spectrum \citep{CotiZelati2021}, as its bright radio counterpart was clearly detected with MeerKAT at 1.28\,GHz but not detected with ATCA at 5.5 or 9\,GHz, although those observations were not taken simultaneously so the difference could simply be a result of source changes (e.g., the long-term variability discussed in Section~\ref{sec:long_term_var}).

What emission mechanism can produce such an exceptionally steep spectrum? The sources that immediately come to mind are radio pulsars, which have an average spectral index of $\alpha\sim-1.6$ over the entire population. Millisecond pulsars of the eclipsing ``Redback'' class \citep{Roberts2013}, to which all 3 known tMSPs belong, have shown even steeper spectral indices \citep{archibald2009,Broderick2016}.
Given that there are multiple pieces of evidence that the radio pulsar of \jttt\ is still active during its low luminosity LMXBs-like state \citep{CotiZelati2014,Jaodand2016} and given the many similarities between \fgl\ and \jttt, it is a reasonable assumption that the radio pulsar can be active in the LMXB state of \fgl\ too. Dedicated low-frequency observations of \fgl\ as well as pulsar search observations could test this hypothesis.

Following the suggested explanation of \citet{Bogdanov2018} for the anti-correlated radio and X-ray emission of \jttt, which involves an active radio pulsar rather than a jet-like outflow (where increased radio flux is attributed to a rapid plasma discharge from the pulsar's magnetosphere or a short-lived pulsar wind nebula), a similar mechanism could explain the anti-correlated X-ray and radio emission of \fgl\ in its low X-ray modes. The tentative finding of the very steep radio spectrum of \fgl\ during these low X-ray modes further supports this hypothesis. However, due to the low statistical significance of our results, more extensive observational campaigns with improved instantaneous sensitivity (e.g. next-generation VLA) are needed to explore this in greater detail.

\subsection{\lr/\lx\ of \fgl\ in comparison with other tMSPs and tMSP candidates}

The radio and X-ray luminosities of \fgl\ occupy the same area of the \lr/\lx\ diagram as confirmed tMSPs (see Figure~\ref{fig:lrlx}). Its position on the diagram is the closest to \jttt\ between other sources; in fact, the lower-limit of the distance (shown by the diagonal dotted lines for each point) puts \fgl\ virtually on top of the \jttt\ points, signifying remarkable similarities between the two sources. 

\fgl\ is also the faintest radio source between all of the tMSP candidates during their low-luminosity LXMB-like state, even after taking into account the uncertainties in the distances of these sources. The only other radio faint tMSP candidate is CXOU~J110926.4$-$65022 \citep{CotiZelati2021}: its bright 1.28\,GHz radio luminosity measurements are shown on Figure~\ref{fig:lrlx}, while the source was undetected at 5.5 or 9\,GHz (the band at which the majority of the \lr/\lx\ measurements are presented in this diagram). The 18\,$\mu$Jy ATCA upper-limit would put CXOU~J110926.4$-$65022 at the same level as a high X-ray mode measurement of \fgl. 

The other two candidate tMSPs 3FGL~J0427.8$-$6704 \citep{Li2020} and NGC~6652B \citep{Paduano2021} show brighter radio and X-ray emission than \fgl. This is not surprising because, in contrast to \fgl, CXOU~J110926.4$-$65022 and \jttt, they reside in the so-called X-ray flare-dominated mode, where the radio and X-ray emission could have quite different origins. Indeed, as concluded by \citet{Li2020}, the radio emission of 3FGL~J0427.8$-$6704 is consistent with a partially self-absorbed compact jet (commonly observed in BH and NS-LMXBs) due to its flat spectral index and lack of short-term radio variability, while a compact-jet origin of \jttt's radio emission is challenging (see Section~\ref{sec:into-rx_tmsps}).

These new measurements of \fgl\ confirm that tMSPs tend to be radio brighter than other classes of NS-LMXBs at similarly low X-ray luminosities. As indicated by correlation analysis of the entire population of NS-LMXB \lr/\lx\ measurements, tMSPs do not match the same power-law relation as tentatively found for other NSs (\lr$\propto$\lx$^{0.44}$; \citealt{Gallo2018}). This is not surprising, given that radio emission of tMSPs is likely of a different origin than the classic partially self-absorbed compact jet for which this relation holds. However, \lr\ measurements of both \jttt\ and \fgl\ in the high X-ray mode are consistent with the \lr\ upper-limits of other NS-LXMBs. This agrees well with the scenario proposed by \citet{Baglio2023} that the radio emission of \jttt\ during its high X-ray mode comes from an optically thick jet. If future measurements confirm this hypothesis, perhaps \lr/\lx\ measurements of tMSPs could indeed be combined with the rest of the NS population via careful selection of high X-ray mode data associated with flat-spectrum parts of the radio lightcurve. This can help to improve our understanding of the relation between the power of the radio jet and the mass accretion rate in neutron stars, since tMSPs reside for prolonged periods of time (from a few weeks up to multiple years; e.g. \citealt{Papitto2013}) in a convenient intermediate X-ray luminosity range (\lx$\sim10^{32-34}$erg s$^-1$) that is difficult to probe in other types of NS-LMXBs (that typically spend no longer than a few days days in that range on their way into or out of their bright outbursts). 

\section{Conclusions}

\begin{itemize}
\item  We detected variable radio emission from \fgl, which, in the case of the second epoch, seems to be in anti-correlation with the low X-ray fluxes of the source. Notably, this behaviour is unique among tMSP candidates and resembles that of \jttt.

\item Significant differences in radio behaviours of \fgl\ were identified between the two observational epochs of our campaign. These differences could be attributed to long-term variability of the source, its X-ray moding behaviour, or a potentially steep radio spectrum of \fgl's radio emission.

\item Radio and X-ray luminosities of \fgl\ during its low and high X-ray modes well match those of \jttt\ and XSS~J12270$-$4859 in the low-luminosity LMXB state. Other tMSP candidates have higher then \fgl's \lr\ for a given \lx. \fgl\ is the most similar to \jttt\ between other candidates.

\item The low radio and X-ray fluxes of the source are the biggest limitation in studying the short-term variability of its emission. No current radio facility is capable of capturing similar behavior to that seen in \jttt, in systems that are $\sim$3 times further away. This will require future facilities such as the next generation VLA and Square Kilometre Array.
\end{itemize}

\section*{Acknowledgements}

The National Radio Astronomy Observatory is a facility of the National Science Foundation operated under cooperative agreement by Associated Universities, Inc. The scientific results reported in this article are based in part on observations made by the Chandra X-ray Observatory, which is operated by the Smithsonian
Astrophysical Observatory for and on behalf of NASA under contract NAS8-03060. This research has relied on software provided by the Chandra X-ray Center (CXC) in the application packages CIAO and software provided by the High Energy Astrophysics Science Archive Research Center (HEASARC), which is a service of the Astrophysics Science Division at NASA/GSFC and the High Energy Astrophysics Division of the Smithsonian Astrophysical Observatory. We acknowledge extensive use of NASA’s Astrophysics Data System Bibliographic Services and arXiv. The Dunlap Institute is funded through an endowment established by the David Dunlap family and the University of Toronto. NG acknowledges partial funding from Nederlandse Onderzoekschool Voor Astronomie (NOVA) and the Natural Sciences and Engineering Research Council of Canada (NSERC; ). AP was partially supported by the grant PID2021-124581OB-I00 funded by MCIU/AEI/10.13039/501100011033 and 2021SGR00426 of the Generalitat de Catalunya. ADJ was partially supported by the Chandra X-ray Center Award No. GO0-21067X.

\section*{Data Availability}
All data analysed in this article are public. The VLA data can be downloaded from the NRAO Data Archive \url{https://data.nrao.edu} under project code \texttt{VLA 18A-398}.  The {\it Chandra} data can be downloaded from the Chandra Data Archive \url{https://cda.harvard.edu/chaser/} using observation IDs \texttt{20902 and 20903}; as well as the High Energy Astrophysics Science Archive Research Center (HEASARC) archive \url{https://heasarc.gsfc.nasa.gov/docs/archive.html}. The derived data generated in this research will be shared on reasonable request to the corresponding author.

\bibliographystyle{mnras}
\bibliography{allbib}

\onecolumn
\newpage
\appendix

\section{Radio variability analysis}\label{sec:appedix_var}

\begin{figure}
\centering
\includegraphics[width=0.9\textwidth]{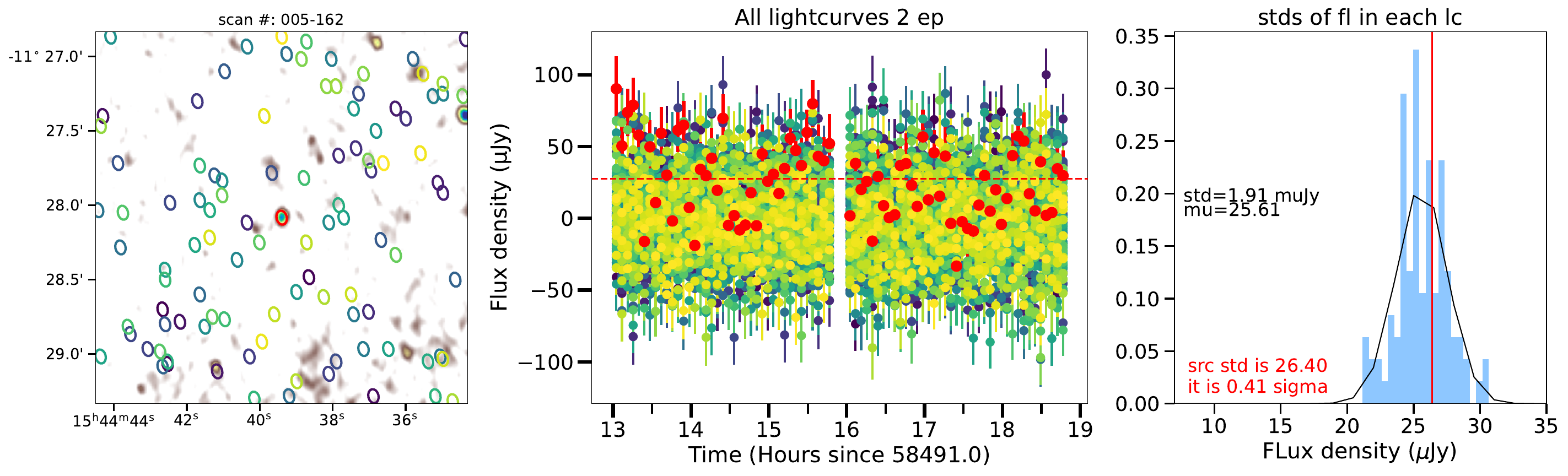}

\caption{Background variability compared to the variability of \fgl\ during epoch 2.
{\it Left}: Radio image of \fgl\, and its surrounding source-free field. Each ellipse shows a random point in the image from which the radio flux density was extracted. {\it Middle}: Radio lightcurves extracted from random locations of the background around \fgl. Red points show the radio lightcurve of \fgl, and the red dashed line represents the average radio flux density of the source. {\it Right}: Distributions of standard deviation of radio flux densities of each background lightcurve compared to that of \fgl\ (shown in red line). This figure shows that the variability of the source is consistent with background variability within 1-$\sigma$.  }\label{fig:background_stat}

\end{figure}

\begin{figure}
\centering
\includegraphics[width=0.9\textwidth]{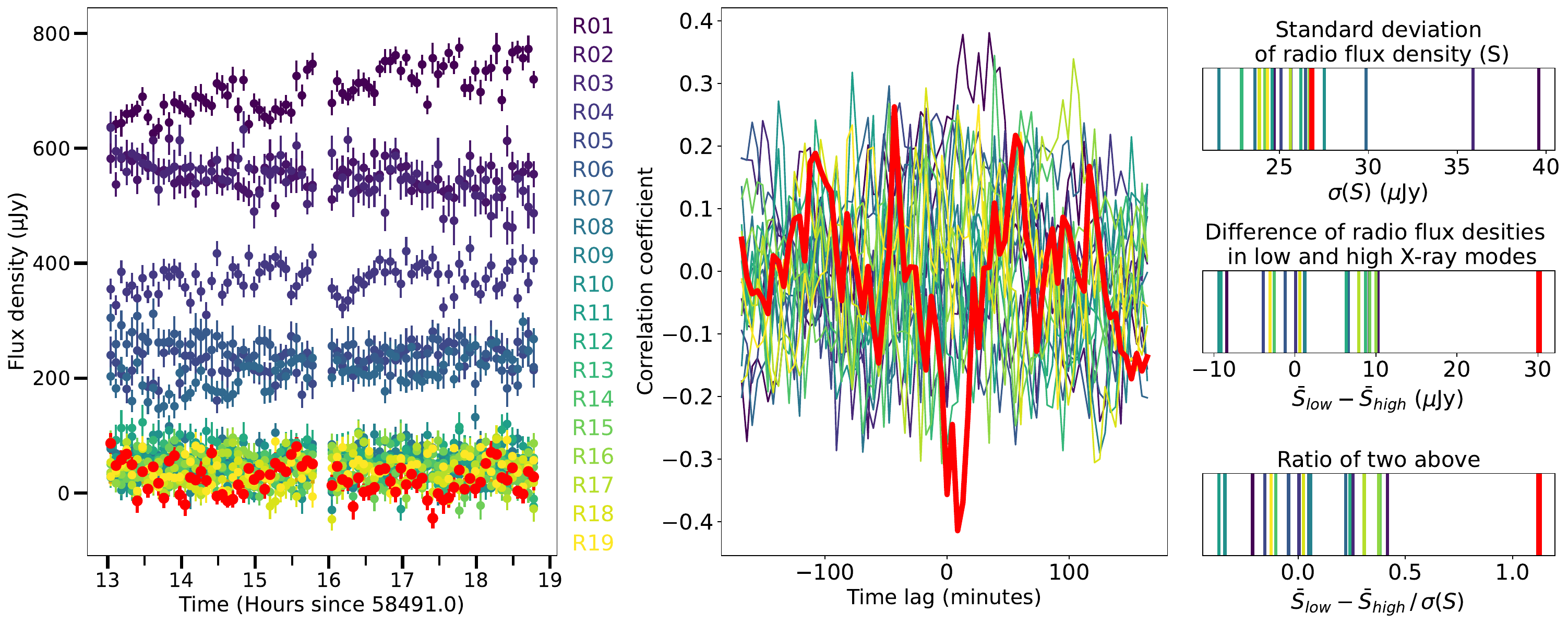}

\caption{Variability comparison of \fgl\ with other sources in the field. {\it Left}: Radio lightcurves of sources R1$-$R19 (shown in Figure~\ref{fig:radio_images}). {\it Middle}: Correlation coefficient for each radio source lightcurve with the X-ray lightcurve of \fgl. The correlation of \fgl\ radio lightcurve is shown by the thick red line. {\it Right}: Comparison of variability properties of all radio sources in the field. From top to bottom: standard deviation of radio flux densities of each lightcurve; the difference of mean of radio flux densities of points that happen to align with the low X-ray modes and the rest of lightcurve of \fgl\ (see top right panel of Figure~\ref{fig:r_vs_x_prob}); the ratio of these two values. It is evident that even though \fgl's variability seems to be consistent with variability of other sources in the field (and even the background, see Figure~\ref{fig:background_stat}), this variability best anti-correlates with the X-ray moding of \fgl, as demonstrated by the fact that it has highest value of correlation coefficient at zero lag, and that its radio flux density is systematically higher in the low X-ray mode compared to the rest of observation. The latter is most evident when comparing with other sources in the field. }\label{fig:variability}
   
\end{figure}

To access the significance of the radio variability observed in lightcurves of \fgl\ obtained from radio imaging of VLA observations, we ran an analysis of short-timescale variability of the noise of our images. Figures~\ref{fig:background_stat} and \ref{fig:variability} provide a summary of this analysis for the second observational epoch. Our findings indicate that the amplitude of \fgl's variability is consistent with that of the background noise (within 1-$\sigma$). However, the particular pattern, associated with the X-ray moding of the system, is unique to radio emission of \fgl, as other sources exhibit no correlation with \fgl's X-ray flux.

%%%%%%%%%%%%%%%%%%%%%%%%%
%%%%%%%%%%%%%%%%%%%%%%%%%

%%%%%%%%%%%%%%%%%%%%%%%%%
%%%%%%%%%%%%%%%%%%%%%%%%%

The middle panel of Figure~\ref{fig:variability} displays the results of correlation of the X-ray and radio lightcurves of \fgl\ (shown in red) and 19 sources from the nearby field. 

\begin{figure*}
\centering
\includegraphics[width=0.9\textwidth]{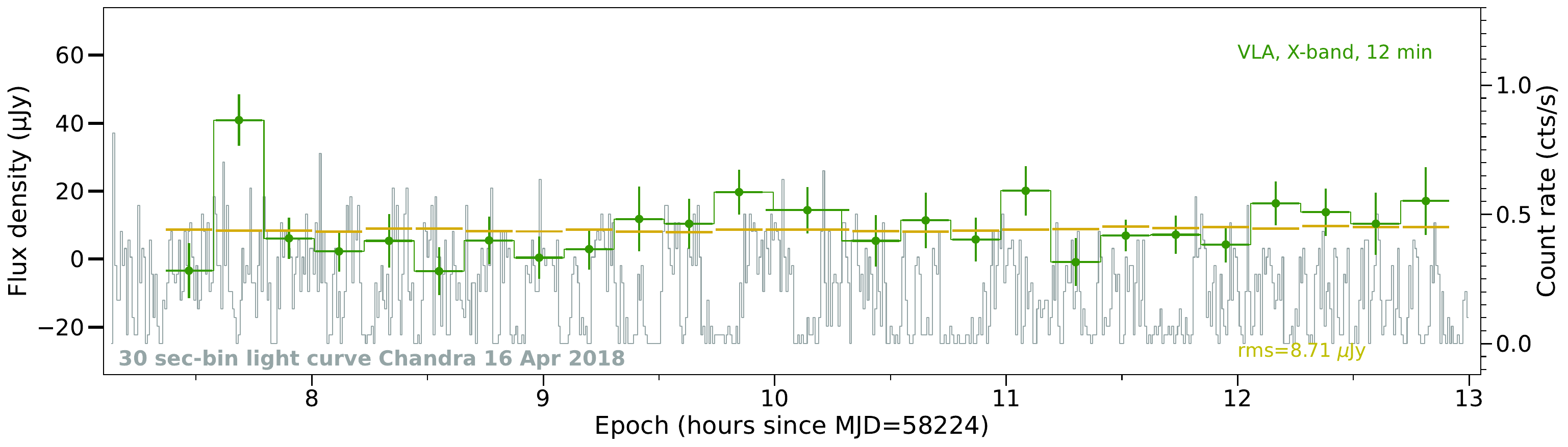}
\includegraphics[width=0.9\textwidth]{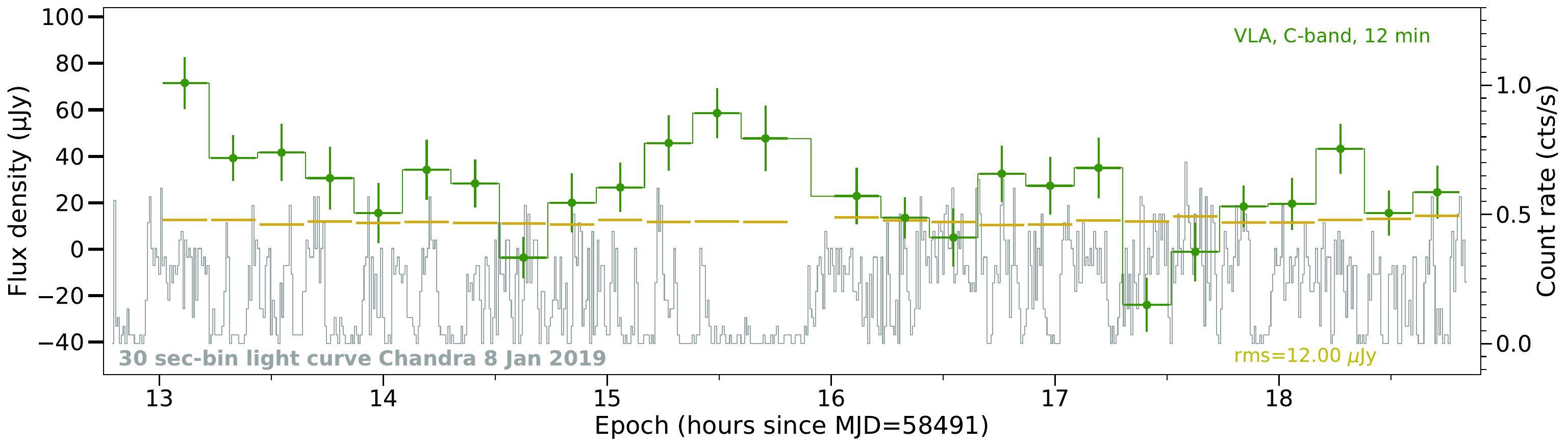}

\includegraphics[width=0.45\textwidth]{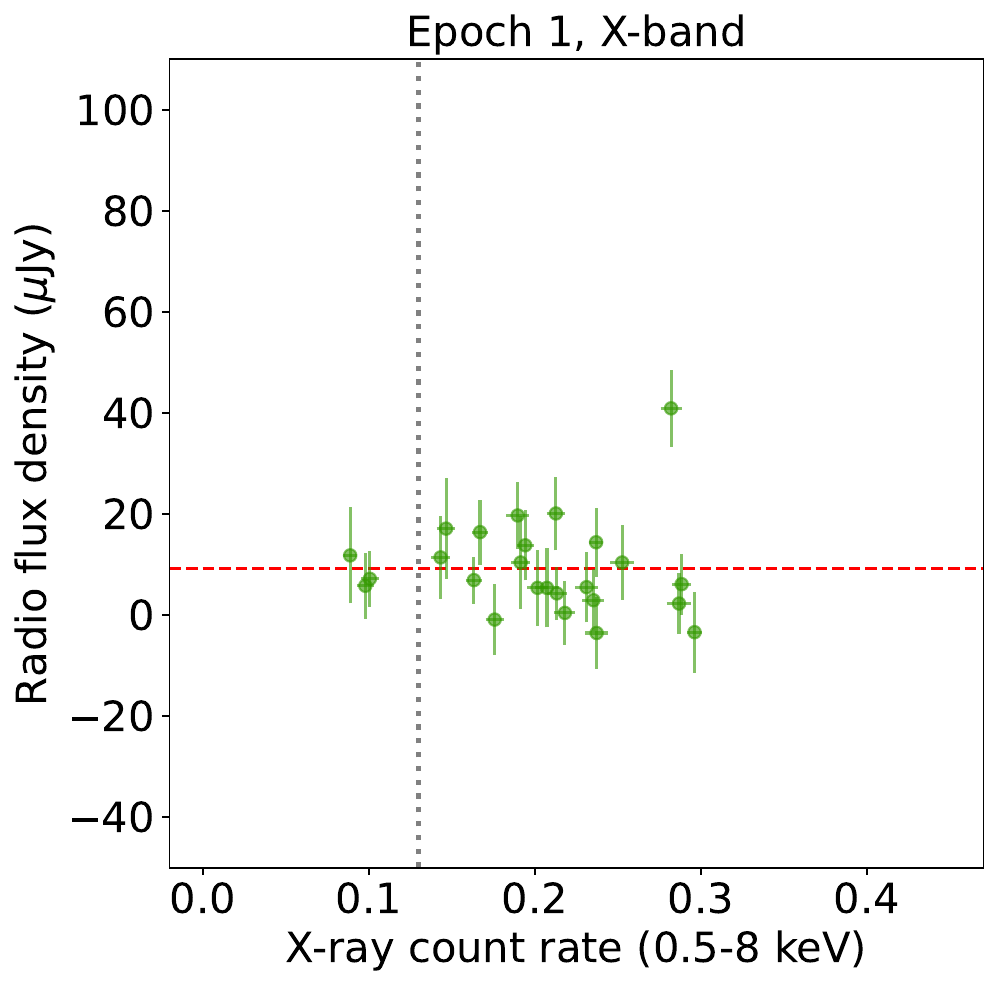}\hspace{10mm}
\includegraphics[width=0.45\textwidth]{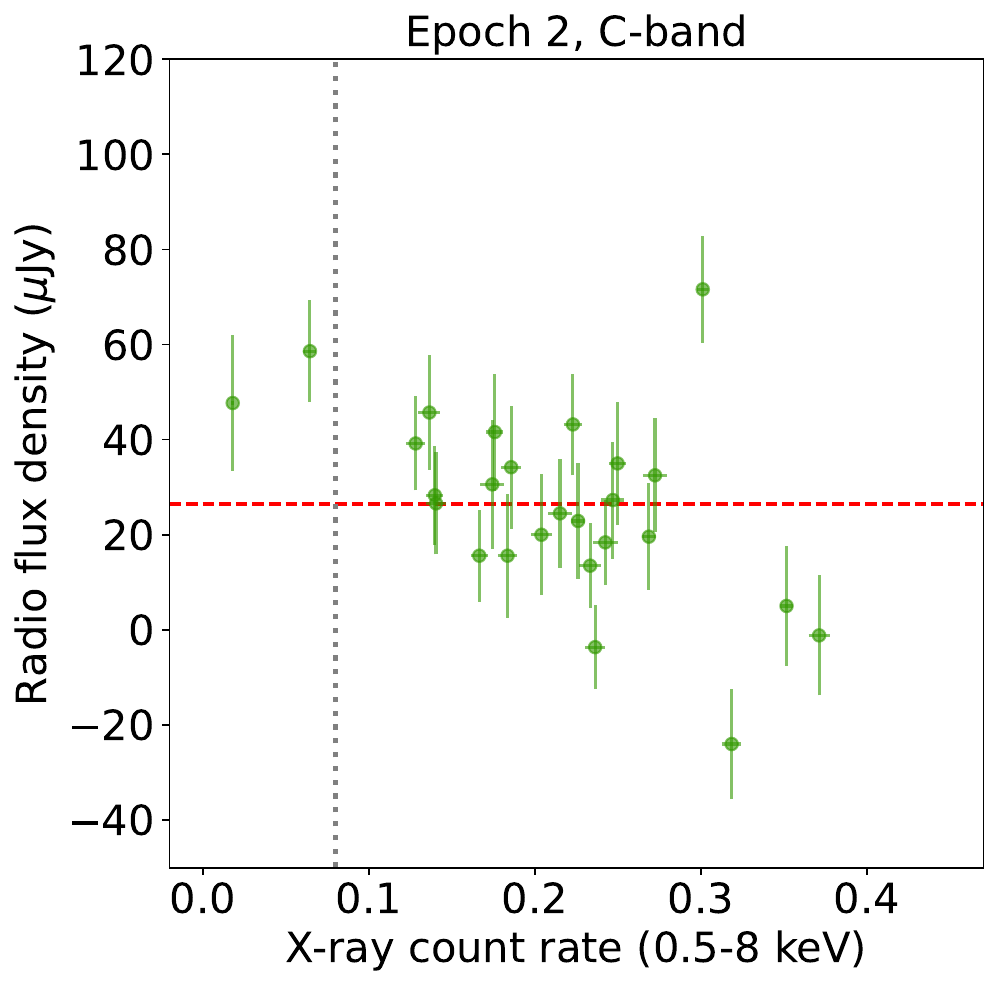}

\caption{{\it Top:} Radio (VLA) and X-ray ({\it Chandra}; $0.5-8$\,keV) lightcurves of \fgl\, from Epoch~1 (top) and Epoch~2 (bottom). Radio lightcurves binned at 12\,min that correspond to averaging of 3 continuous scans of the source. {\it Bottom:} Scatter plot of radio flux density measurements with the corresponding average X-ray count rates done for each of radio lightcurves from panels above, shown for Epoch~1 ({\it left}) and Epoch~2 ({\it right}). }\label{fig:3sc_lcs_scaller}
\end{figure*}

We also present radio lightcurves produced using larger time-bin, averaging 3-scans of the source (see Figure~\ref{fig:3sc_lcs_scaller}, top two panels). In the second epoch, the source appears to be detected in a larger number of radio bins. Despite the longer averaging, leading to more X-ray mode confusion for each radio bin, the anti-correlation between radio and X-ray fluxes of \fgl\ could still be visible in the second epoch (see Figure~\ref{fig:3sc_lcs_scaller}, bottom right panel). Additionally, a distinct bright radio flare could be seen in both epoch, occurring at the corresponding high X-ray count rates. This may serve as evidence for a similar ``orphan" flaring that is observed in \jttt\ simultaneous {\it Chandra} and VLA observation \citep{Bogdanov2018}.

% Don't change these lines
\bsp	% typesetting comment
\label{lastpage}
\end{document}